\documentclass[aps,prl,reprint,twocolumn]{revtex4-2}

\usepackage[utf8]{inputenc}
\usepackage[T1]{fontenc}
\usepackage{mathptmx}
\usepackage{etoolbox}
\usepackage{graphicx}

\usepackage{hyperref}
\usepackage{orcidlink}
\usepackage[scaled=0.9]{helvet}
\usepackage{courier}
\usepackage[letterpaper,top=0.88in,bottom=0.87in,
            left=0.68in,right=0.68in,
            columnsep=0.25in]{geometry}

\usepackage{lmodern}
\rmfamily
\DeclareFontShape{T1}{lmr}{b}{sc}{<->ssub*cmr/bx/sc}{}
\DeclareFontShape{T1}{lmr}{bx}{sc}{<->ssub*cmr/bx/sc}{}

\usepackage{amsmath,amssymb}

\usepackage{mathtools,nccmath}

\usepackage{bm}
\usepackage{setspace}
\usepackage{flushend}

\usepackage{xargs}
\usepackage[usestackEOL]{stackengine}
\usepackage{paralist}

\usepackage{xstring}
\usepackage{tikz}
\usepackage{scalerel}

\usepackage{pgffor}

\definecolor{myblue}{rgb}{0.0, 0.0, 0.85}
\hypersetup{
    colorlinks=true,
    linkcolor=myblue,
    filecolor=magenta,
    urlcolor=myblue,
    citecolor=myblue
}
\usepackage[capitalise,nameinlink]{cleveref}
\crefname{section}{Sec.}{Secs.}
\Crefname{section}{Sec.}{Secs.}

\usepackage{xfrac}

\usepackage{wasysym}

\newcommand{\pp}{\partial}

\newcommand{\dd}{\mathrm{d}}

\newcommand*{\IsInteger}[3]{%
    \IfStrEq{#1}{ }{%
        #3%
    }{%
        \IfInteger{#1}{#2}{#3}%
    }%
}%

    \newcommandx{\indexSequence}[4][1=1,2=4,3={},4=\alpha]{
        {
        \ifnum\the\numexpr#2-#1\relax>-1
            \ifnum #2=#1
                #3{#4_{#1}}
            \else
                \ifnum\the\numexpr#2-#1\relax<3
                    \foreach \i in {#1,...,#2}
                    {
                    #3{#4_{\i}}
                    }
                \else
                    {
                    #3{#4_{#1}}..#3{#4_{#2}}
                    }
                \fi
            \fi
        \else
            #3{#4_{\text{error (#2<#1)}}}
        \fi
        }
    }
    \newcommandx{\Tensor}[6][1=T,2=1,3=1,4={},5=\alpha,6={}]{
        \def\literal{#1_{#6#4{#5_{#2}}..#4{#5_{#3}}}}
        \IsInteger{#2}{
            \IsInteger{#3}{
                #1_{#6\indexSequence[#2][#3][#4][#5]}
                }{
                \literal
            }
        }{
            \literal
        }
    }
    \newcommandx{\ifstrnotempty}[3]{\ifstrempty{#1}{#3}{#2}}
    \newcommandx{\sIndex}[2][2=\alpha]{\IsInteger{#1}{#2_{#1}}{#1}}
    \newcommandx{\MTensor}[6][1=T,2=1,3={},4={},5={},6={}]{
        \ifstrnotempty{#6}{
            #1_{{\sIndex{#2}}{\sIndex{#3}}{\sIndex{#4}}{\sIndex{#5}}{\sIndex{#6}}}
        }
        {
            \ifstrnotempty{#5}{
                #1_{{\sIndex{#2}}{\sIndex{#3}}{\sIndex{#4}}{\sIndex{#5}}}
            }
            {
                \ifstrnotempty{#4}{
                    #1_{{\sIndex{#2}}{\sIndex{#3}}{\sIndex{#4}}}
                }
                {
                    \ifstrnotempty{#3}{
                        #1_{{\sIndex{#2}}{\sIndex{#3}}}
                    }
                    {
                        #1_{{\sIndex{#2}}}
                    }
                }
            }
        }
    }

    \let\T\Tensor
    \let\MT\MTensor

\newcommand\Wtilde[1]{\ThisStyle{%
  \setbox0=\hbox{$\SavedStyle#1$}%
  \stackengine{-.1\LMpt}{$\SavedStyle#1$}{%
    \stretchto{\scaleto{\SavedStyle\mkern.2mu\AC}{.5150\wd0}}{.3\ht0}%
  }{O}{c}{F}{T}{S}%
}}

\newcommand{\zero}{{(0)}}
\newcommand{\one}{{(1)}}
\newcommand{\two}{{(2)}}

\newcommand{\Neq}{{\rm neq}}

\newcommand{\sgs}{{\rm sgs}}

\renewcommand\Re{\mathrm{Re}}
\newcommand\Ma{\mathrm{Ma}}

\newcommand\Kn{\mathrm{Kn}}

\newenvironment{alignedEq}[1][\unskip]{\equation #1 \aligned}{\endaligned\endequation}
\newenvironment{alignedEq*}[1][\unskip]{\begin{equation*} #1 \aligned}{\endaligned\end{equation*}}
\newenvironment{alignatEq}[2][\unskip]{\equation #1 \alignedat{ #2 }}{\endalignedat\endequation}

\makeatletter
\def\@email#1#2{%
 \endgroup
 \patchcmd{\titleblock@produce}
  {\frontmatter@RRAPformat}
  {\frontmatter@RRAPformat{\produce@RRAP{*#1\href{mailto:#2}{#2}}}\frontmatter@RRAPformat}
  {}{}
}%

\def\underbracex#1#2{\mathop{\vtop{\m@th\ialign{##\crcr
   $\hfil\displaystyle{#2}\hfil$\crcr
   \noalign{\kern3\p@\nointerlineskip}%
   #1\crcr\noalign{\kern3\p@}}}}\limits}

\def\underbracea{\underbracex\upbracefilla}

\def\upbracefilla{$\m@th \setbox\z@\hbox{$\braceld$}%
  \bracelu\leaders\vrule \@height\ht\z@ \@depth\z@\hfill
\kern\p@\vrule \@width\p@\kern\p@\vrule \@width\p@\kern\p@\vrule \@width\p@
$}

\def\upbracefillb{$\m@th \setbox\z@\hbox{$\braceld$}%
\vrule \@width\p@\kern\p@\vrule \@width\p@\kern\p@\vrule \@width\p@\kern\p@
 \leaders\vrule \@height\ht\z@ \@depth\z@\hfill\bracerd
  \braceld\leaders\vrule \@height\ht\z@ \@depth\z@\hfill
\kern\p@\vrule \@width\p@\kern\p@\vrule \@width\p@\kern\p@\vrule \@width\p@
$}

\def\upbracefillc{$\m@th \setbox\z@\hbox{$\braceld$}%
\vrule \@width\p@\kern\p@\vrule \@width\p@\kern\p@\vrule \@width\p@\kern\p@
\leaders\vrule \@height\ht\z@ \@depth\z@\hfill
\kern\p@\vrule \@width\p@\kern\p@\vrule \@width\p@\kern\p@\vrule \@width\p@
$}

\def\upbracefilld{$\m@th \setbox\z@\hbox{$\braceld$}%
\vrule \@width\p@\kern\p@\vrule \@width\p@\kern\p@\vrule \@width\p@\kern\p@
 \leaders\vrule \@height\ht\z@ \@depth\z@\hfill\braceru$}

\def\underbracebd{\underbracex\upbracefillbd}
\def\upbracefillbd{$\m@th \setbox\z@\hbox{$\braceld$}%
\vrule \@width\p@\kern\p@\vrule \@width\p@\kern\p@\vrule \@width\p@\kern\p@
\bracerd\braceld
 \leaders\vrule \@height\ht\z@ \@depth\z@\hfill\braceru$}
\newcommand{\pushright}[1]{\ifmeasuring@#1\else\omit\hfill$\displaystyle{#1}$\fi\ignorespaces}
\newcommand{\pushleft}[1]{\ifmeasuring@#1\else\omit$\displaystyle{#1}$\hfill\fi\ignorespaces}

\makeatother

\begin{document}

\title{Kinetic closure of turbulence}
\author{Francesco Marson$^{1,2,*}$\,\orcidlink{0000-0002-0900-4193} and Orestis Malaspinas$^{1}$\,\orcidlink{0000-0001-9427-6849}}
\email{francesco.marson@proton.me}
\affiliation{%
  $^{1}$HEPIA, University of Applied Sciences and Arts of Western Switzerland%
}
\affiliation{$^{2}$Department of Computer Science, University of Geneva}%

\date{September 23, 2026}

\begin{abstract}
This letter presents a kinetic closure of the filtered Boltzmann--BGK equation, paving the way toward an alternative description of turbulence. The closure retains the turbulent subfilter stress tensor without a separate Smagorinsky-type ansatz for its structure, unlike classical filtered Navier--Stokes closures. In contrast, it accounts for the subfilter turbulent diffusion in the nonconserved moments by generalizing the BGK collision operator. The model does not require scale separation between resolved and unresolved scales. The Chapman--Enskog analysis shows how its hydrodynamic limit can converge to the filtered Navier--Stokes equations, with velocity gradients isolating subfilter contributions. Numerical tests on the Taylor--Green vortex and the turbulent mixing layer show improved stability and reduced dissipation in the reported cases, benchmarked against the Smagorinsky model.
\end{abstract}

\maketitle

\newcommand{\F}{\mathcal{F}}

Turbulent flows are ubiquitous in nature, yet their chaotic and multiscale dynamics make them notoriously difficult to model.
This complex character poses a primary challenge for numerically solving their governing equations. In principle, direct numerical simulation (DNS) could resolve all scales, but in practice, it remains computationally intractable for most flows of engineering interest~\cite{pope_turbulent_2000,davidson_turbulence_2015}.
A common remedy is to apply the governing equations to filtered rather than local flow variables~\cite{reynolds_iv_1895,smagorinsky_general_1963,lilly_representation_1966}.
Consider the generic transport equation for a filtered scalar $\bar\phi$ advected by the filtered velocity $\bar{\bm{u}}$:
\begin{equation}\label{transport}
    \pp_t \bar\phi + T(\bar{\bm{u}}, \bar{\phi}) + \mathcal{E}_T(\bm u,\bm {u}\phi,\phi)
 = D(\bar{\bm{u}}, \bar{\phi}) + \mathcal{E}_D(\bm u,\bm {u}\phi,\phi),
\end{equation}
where the filtered counterpart of any quantity \(a\) is defined as
\(\bar a=\mathcal{G}_{\Delta}\circ a\), with \(\mathcal{G}_{\Delta}\) being a convolution operator satisfying the standard properties of large-eddy simulation (LES) filters: conservation of constants, linearity, and commutation with derivatives~\cite{smagorinsky_general_1963,sagaut_large_2006}.
In \cref{transport}, the hyperbolic operator \(T\) represents advective fluxes of \(\bar\phi\), while the parabolic operator \(D\) accounts for its diffusive transport, which drives entropy production.

\begin{figure}
    \centering
    \includegraphics[scale=0.9]{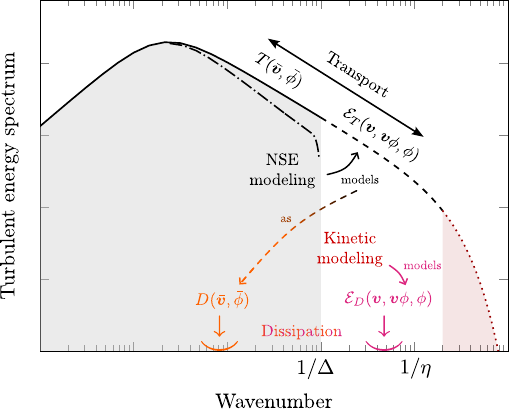}
    \caption{Turbulent energy spectrum and the general transport equation terms.}
    \label{fig:turbulent}
\end{figure}

When coarse-graining the flow dynamics as in \cref{transport}, filtering residuals ($\mathcal{E}$) arise from the dependence on unresolved scales~\cite{sagaut_large_2006}.
These terms represent the influence of filtered turbulent fluctuations and fall into two categories:
(i) the diffusive residual $\mathcal{E}_D$, which modifies the dissipation of $\phi$ into thermal energy at subfilter scales; and
(ii) the convective residual $\mathcal{E}_T$, which alters the transport of $\phi$ in physical space and across scales.
The latter links to the Kolmogorov cascade~\cite{kolmogorov_local_1941}, where large eddies break down into smaller subfilter-scale vortices, and to the inverse process of backscatter, which transfers momentum from subfilter to resolved scales~\cite{sagaut_large_2006}.
\Cref{fig:turbulent} illustrates the connection of these terms with the energy spectrum.

In the filtered Navier--Stokes equation (NSE) commonly used in LES~\cite{navier_memoire_1822,stokes_theories_1845,smagorinsky_general_1963,lilly_representation_1966}, the dominant nonlinear filtering residual is the subfilter-scale stress tensor~\cite{smagorinsky_general_1963,lilly_representation_1966,sagaut_large_2006}. This term is convective, corresponds to $\mathcal E_T$ in \cref{transport}, and is typically modeled using a turbulent viscosity model~\cite{davidson_turbulence_2015} (e.g., amounting to a turbulent dissipation; see \cref{fig:turbulent}). The diffusive residual $\mathcal E_D$ is absent from the filtered NSE due to the linearity of the viscous stress tensor expressed through Newton's constitutive laws. This modeling approach is justified because $\mathcal E_T$ primarily accounts for the transport of momentum from resolved to unresolved scales; once transferred to the unresolved scales, momentum is dissipated in dynamics that are not governed by the equation above but by higher-order additional equations, such as those used in Reynolds-averaged Navier--Stokes (RANS) modeling \cite{rotta_statistische_1951,pope_turbulent_2000,ferziger2020computational}.

The Boltzmann--BGK equation (BGK--BE) is a kinetic equation that can also describe flow dynamics. It follows from the Boltzmann equation (BE) \cite{boltzmann_weirere_1872} by using the BGK collision model \cite{bhatnagar_model_1954}. In contrast to the NSE, the linear nature of the BGK--BE transport terms prevents a convective filtering residual $\mathcal E_T$. Therefore, the only filtering residual to be modeled in the BGK--BE appears in the collision model term, which is diffusive and corresponds to $\mathcal E_D$~\cite{ansumali_kinetic_2004,malaspinas_consistent_2012} (see \cref{fig:turbulent}).

We argue that this fundamental dissimilarity makes the kinetic closure (KC) of turbulence intrinsically different from the NSE case.
Here, we exploit this distinction to introduce a kinetic closure model for turbulent flows.
To this end, we depart from the conventional assumptions that have guided earlier kinetic approaches to turbulence modeling, namely:
(i) the renormalization of the collision operator to model subgrid effects purely via an effective relaxation time \cite{chen_analysis_1999, succi_towards_2002, chen_extended_2003};
(ii) the reliance on macroscopic eddy-viscosity concepts \cite{girimaji_boltzmann_2007, banda_discrete-velocity_2006} or structural closures that exploit specific mathematical properties of the spatial filter to reconstruct subgrid stresses \cite{sagaut_toward_2010,malaspinas_advanced_2011,ansumali_kinetic_2004};
(iii) the use of perturbative expansions to enforce consistent scaling between the filter width and the Knudsen number, a procedure that inherently restricts validity to regimes of spectral scale separation \cite{ansumali_kinetic_2004,luan_constructing_2025};
(iv) phenomenological descriptions that rely on thermodynamic analogies to construct equilibrium distributions \cite{chen_expanded_2004, chen_non-perturbative_2013, chen_average_2023, saveliev_kinetic_2024,luan_constructing_2025}; and
(v) the coupling with auxiliary macroscopic transport equations (e.g., $k-\varepsilon$) to determine relaxation parameters \cite{asinari_kinetic_2016,righi_gas-kinetic_2016}.
Finally, we distinguish our pragmatic closure from classical theoretical attempts to close the BBGKY kinetic hierarchy by leveraging higher-order nonlocal correlations \cite{zhigulev_equations_1965,tsuge_approach_1974,chliamovitch_truncation_2015, chliamovitch_kinetic_2017}.
While theoretically rigorous, such formulations generally yield high-dimensional systems that are ill-suited for practical engineering applications. By moving beyond these heuristic, macroscopic, or purely theoretical constraints, we aim to establish a self-consistent and operational kinetic theory for turbulent flows.

\newcommand{\cpartial}{\pp^{\star}}
\newcommand\Lcc{\mathcal{L}}
\newcommand\Tcc{\mathcal{T}}
\newcommand\Ucc{\mathcal{U}}
Let us consider the nondimensional BGK--BE, obtained by scaling with the Mach and Reynolds numbers ($\Ma$ and $\Re_{\ell}$).
Einstein's summation convention will be assumed throughout the remainder of this letter.
The resulting nondimensional BGK--BE is given by:
\begin{equation}
    \label{eq:BEdimless2}{
    \cpartial_t f + \xi^{\star}_{\alpha}\cpartial_{\alpha}f=-\frac{\Re_{\ell}}{{\Ma^2}}\,{\omega^*}\left({f-f^\zero}\right)\,,
    }
\end{equation}
with the following nondimensional quantities:
    {\allowdisplaybreaks
    \begin{alignat*}{2}
        &\cpartial_t \equiv \Tcc\partial_t,\   \cpartial_\alpha \equiv \Lcc\partial_\alpha &\text{(time/space derivatives)}\\
        &f \equiv f^\star \equiv \Ucc^3 m \cdot n / m_{\rm ref}\quad &\text{(mass PDF)}\\
        &\xi_\alpha^\star \equiv \sfrac{\xi_\alpha}{\Ucc},\ \mathrlap{\zeta_\alpha^*\equiv \sfrac{(\xi_\alpha-u_\alpha)}{\sqrt{\theta_R}}}\ &\text{(particle/peculiar vel.)}\\
        &\Re_\ell \equiv \sfrac{\Ucc\Lcc}{ \sqrt{\theta_R}\ell},\  \Ma\equiv\sfrac{\Ucc}{\sqrt{\theta_R}}\  &\text{(Reynolds/Mach)}\\
        &\omega^* \equiv \omega\,\ell/\sqrt{\theta_R}\quad &\text{(collision frequency)}\\
        &f^\zero \equiv\frac{\rho\Ma^3}{(2\pi \theta^*)^\frac{3}{2}}e^{-\frac{\zeta^{*2}}{2\theta^*}}
        \ &\text{(Maxwellian)}\\
        &\rho \equiv \int_\Xi f \, \dd\bm\xi^\star,\ \mathrlap{u_\alpha^\star \equiv \int_\Xi f \xi_\alpha^{\star} \, \dd \bm\xi^{\star}/\rho}\quad &\text{(density/velocity)}\\
        &\theta^* \equiv \int_\Xi f \zeta_\alpha^{*2} \, \dd\bm\xi^{\star}/(3\rho). &\text{(fluid temperature)}
    \end{alignat*}}
Here, $n$ is the \textit{probability density function} (PDF), $m$ and $m_{\rm ref}$ are respectively the particle and the reference masses, and $\alpha \in \{x,y,z\}$ is the index of the space coordinate.
The nondimensional \(f\) is the corresponding local convective velocity-space density, written with respect to \(d\bm\xi^\star\).
If \(f^*\) denotes the same distribution written with the diffusive velocity variable \(\bm\xi^*=\bm\xi/\sqrt{\theta_R}\), then \(f\,d\bm\xi^\star=f^*\,d\bm\xi^*\), hence \(f=\Ma^3 f^*\), which clarifies the origin of the \(\Ma^3\) factor in \(f^\zero\).
$\Lcc,\Tcc,\Ucc = \Lcc/\Tcc$ are respectively the convective reference length, time and velocity; $\ell,\sqrt{\theta_R}\equiv  \sqrt{{\int \zeta_\alpha^2 f_R\dd \zeta_\alpha}/{[3\rho(f_R)]}}$  are respectively the diffusive reference length (mean free path of particles) and velocity (square-root of the reference temperature) based on a reference distribution function $f_R$ and the peculiar velocity $\zeta_\alpha \equiv \xi_\alpha - u_\alpha$; $\omega$ is the relaxation frequency; \(\Xi\) is the velocity space; finally, the symbol $*$ denotes the nondimensionalization using $\sqrt{\theta_R}$ and $\ell$, while $\star$ indicates a nondimensionalization of the variable using $\Ucc$ and $\Lcc$.

We now consider the filtering of the BGK--BE.
Applying the filter to the nondimensional BGK--BE in \cref{eq:BEdimless2}, we obtain the filtered BGK--BE (FBGK--BE):
\begin{equation}\label{fbe_bgk}
    \cpartial_t \overline f + \xi^{\star}_{\alpha}\cpartial_{\alpha} \overline f
    = -\frac{\Re_{\ell}}{{\Ma^2}}\omega^*\left({\overline f- \overline{f^{\zero }}}\right)\,.
\end{equation}
In \cref{fbe_bgk}, $\overline{f^{\zero }}$ is not directly computable from $\overline f$, and requires knowledge of $f$.
Therefore, \cref{fbe_bgk} is not closed in $\overline f$, and $\overline{f^{\zero }}$ conceals filtering residuals, which can be made explicit with the usual decomposition:
\begin{equation}\label{f0}
    \overline{f^\zero}
    = \underline f^\zero\left(\overline f\right)+
    f_{\sgs}(f)\,,
\end{equation}
where $\underline f^\zero \equiv f^{(0)}\left(\bar \rho\left(\overline f\right),{\tilde u_\alpha\left(\overline f\right)}, \tilde\theta^*\left(\overline f\right) \right)$, and $f_{\sgs}(f)\equiv \overline{f^\zero}-\underline f^\zero$ is the subfilter-scale (or subgrid-scale, SGS) equilibrium distribution. Here, we used the following Favre--averaged quantities:
\(\MT[\tilde u^\star][1] \equiv \sfrac{\overline{\rho \MT[u^\star][1]}}{\bar \rho }\),
\(\tilde \theta^* \equiv \sfrac{\overline{\rho \theta^*}}{\bar \rho }\,\)~\cite{favre_equations_1965,favre_turbulence_1983}.

The discussion up to this point, specifically in \cref{fbe_bgk,f0}, closely aligns with the initial efforts to model turbulence from a kinetic perspective~\cite{succi_towards_2002}. However, our present approach is to maintain generality in the expansion procedure, in the dimensional analysis, and by
generalizing the BGK collision model in the filtered case.

\begin{figure}
    \centering
    \includegraphics[scale=0.85]{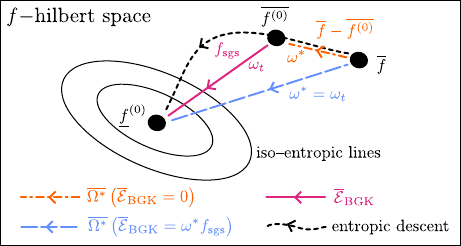}
    \caption{Representation of the Hilbert space of $f$.}
    \label{hilbert}
\end{figure}

The BGK collision model describes the rate of change of \(f\) due to particle collisions, per unit time.
This process is assumed to be linear, Markovian, and much faster than the convective time scale. This justifies modeling the variation of \(f\) as a relaxation toward a fixed point \(f^\zero\), a Maxwell--Boltzmann distribution that can be derived by entropy maximization under fixed macroscopic conserved moments (i.e., density, momentum, and translational kinetic energy).
Let us name \(\mathcal{E}_{\rm BGK}\) the small error associated with these assumptions.
Then, denoting by \(\Omega^*\) the unfiltered collision operator appearing on the right-hand side of \cref{eq:BEdimless2}, one can formally write: \(\Omega^* \equiv -\omega^*(f-f^\zero) - \mathcal{E}_{\rm BGK}\).
Filtering \(\Omega^*\) instead of only \(\omega^*(f-f^\zero)\) yields
\begin{equation}\label{eq:omega_filtered}
    \overline{\Omega^*} \equiv -\omega^*\left(\overline{f}-\overline{f^\zero}\right) - \overline{\mathcal{E}}_{\rm BGK}.
\end{equation}
While neglecting \(\mathcal{E}_{\rm BGK}\) is typically a reasonable approximation, discarding \(\overline{\mathcal{E}}_{\rm BGK}\) \emph{a priori} is not justified.
In \cref{CE} we show with the Chapman--Enskog expansion (CE) that, in fact, \(\mathcal{E}_{\rm BGK}\) is of the same asymptotic order as $\overline f - \overline{f^\zero}$ and plays a role in the dissipation of the subgrid convective terms (see the discussion after \cref{msgs} in \cref{CE}).
In contrast, naively assuming \(\overline{\mathcal{E}}_{\rm BGK} = 0\) corresponds to the relaxation process shown by the dot-dashed orange line in \cref{hilbert}.

The collision term in \cref{fbe_bgk} suffers from a further issue: \(\overline{f^\zero}\) cannot be computed directly from the filtered conserved moments.
The standard remedy is to reduce \cref{eq:omega_filtered} to
\(
    \overline{\Omega^*} \approx -\tilde\omega^*(\overline{f}-{\underline f^\zero})
\)
by implicitly assuming \(\overline{\mathcal{E}}_{\rm BGK} \approx \omega^* f_\sgs\) (dashed blue line in \cref{hilbert}).
This assumption, however, restricts the flexibility of assigning a distinct collision frequency to the relaxation of \(f_\sgs\) toward zero (magenta line in \cref{hilbert}), and effectively enforces the use of \emph{ad hoc} dissipative turbulence models borrowed from NSE closures to stabilize the solution.
A straightforward way to generalize \cref{eq:omega_filtered} with the approximation \(\overline{\mathcal{E}}_{\rm BGK} \approx \omega^* f_\sgs\), while preserving the physical meaning of the collision process, is
\begin{equation}\label{generalized_bgk}
    \overline{\Omega^*} \equiv -\omega^*\left(\overline{f}-\overline{f^\zero}\right) - \underbrace{\omega_tf_\sgs}_{\overline{\mathcal E}_{\rm BGK}}.
\end{equation}
$\overline{\Omega^*}$ defined in \cref{generalized_bgk} is a generalized version of the BGK collision model since it converges to BGK for $\Delta\to 0$ because $\lim_{\Delta \to 0} f_\sgs = 0$.
Therefore, the physically consistent equation to close is
\begin{equation}\label{new_filtered}
    \cpartial_t \overline f + \xi^{\star}_{\alpha}\cpartial_{\alpha} \overline f=-\frac{\Re_{\ell}}{{\Ma^2}}\underbrace{\left[\omega^*(\overline{f}-\overline{f^\zero}) + \omega_t f_\sgs\right]}_{-\overline{\Omega^*}},
\end{equation}
where we used the generalized collision operator defined in \cref{generalized_bgk}.

We now propose a closure of \cref{new_filtered} by deriving explicit forms for \( f_\sgs \) and \( \omega_t \).
First, we apply the Chapman--Enskog (CE) expansion to \cref{new_filtered} \cite{enskog_kinetische_1922,chapman_mathematical_1953}. Unlike earlier studies \cite{chen_expanded_2004}, our analysis does not separate turbulent fluctuations from mean-flow dynamics. Instead, it distinguishes between diffusive and convective processes, consistent with the standard treatment of the BGK--BE.
The complete CE procedure is provided in \cref{CE}; under the constitutive split introduced there, the second-order Hermite approximation gives:
\begin{gather}\label{f1approx1}
    \epsilon\overline{f^\one}\;\approx\;
    -\frac{\Ma^2 w \T[H][1][2]}{2\Re}\,
    \left[
        \bar \rho
        \left(
            \MT[\pp^{(1)}][1]\MT[\tilde u^\star][2]
            + \MT[\pp^{(1)}][2]\MT[\tilde u^\star][1]
        \right)
    \right],
    \\[0.5em] \label{closure1}
    f_\sgs \;\approx\; \overline f - \underline f^\zero - \epsilon \,\overline{f^\one}\,,
\end{gather}
where \(w \equiv w^\star=\Ma^3(2\pi)^{-3/2}\exp\!\left( - \sfrac{\xi^{*2}}{2} \right)\) is the fixed-reference Hermite weight written as a convective velocity-space density and employed in \cref{f1} with $\theta^*_R = 1$, and $\Re \equiv \sfrac{\Re_{\ell}}{\tau^*\tilde \theta^*}$.
Both \cref{f1approx1,closure1} depend only on the moments of \(\overline f\), and therefore constitute the first step toward closing \cref{new_filtered}.
These results demonstrate that:
\begin{inparaenum}[(i)]
    \item the FBGK--BE retains information about the SGS tensor $\MT[m^\sgs][1][2]=\int_\Xi {f_\sgs}\xi_{\alpha_1}^\star\xi_{\alpha_2}^\star\dd\bm\xi^\star=\bar\rho(\Wtilde{\MT[u^\star][1]\MT[u^\star][2]}-\MT[\tilde u^\star][1]\MT[\tilde u^\star][2])$ that also appears in the filtered NSE;
    \item turbulent transport is inherently captured, since the dynamics of \(f_\sgs^\zero\) are naturally described without requiring additional transport equations (as in \(k\)--\(\epsilon\) or RANS models);
    \item no Smagorinsky-type assumption is needed for \(\MT[m^\sgs][1][2]\), as its contribution can be estimated from the velocity gradients.
\end{inparaenum}

The last step is to find an expression for $\omega_t$.
Knowing $f_\sgs^\zero$, one can use dimensional analysis to estimate $\omega_t$ like in the $k-\varepsilon$ model~\cite{launder_numerical_1974}
\begin{gather}
    \nu_t \equiv \frac{\theta_R} {\omega_t}
    \overset{\scriptscriptstyle (a)}{\approx}
    C_\nu^\prime \frac{\MT[m^{\sgs 2}][1][1]}{\pp_t\MT[m^\sgs][1][1]}\overset{\scriptscriptstyle (b)}{\approx} C_\nu \Delta \sqrt{\frac{|\MT[m^\sgs][1][1]|}{2\bar\rho}},\label{turbulent_viscosity}
\end{gather}
where
the constant $C_\nu$ has to be determined experimentally and $\MT[m^\sgs][1][2]$ can be computed from $f_\sgs^\zero$.

The approximations introduced so far can be refined in future work, for example, by leveraging recursive formulas for $\overline{f^\one}$ (see \cite{malaspinas_increasing_2015}) in a multi-relaxation-rate framework \cite{malaspinas_increasing_2015,higuera_lattice_1989}. Nevertheless, this is left for future work, and we validate our model with the simple BGK--lattice Boltzmann method (LBM)~\cite{mcnamara_use_1988,higuera_boltzmann_1989,kruger_lattice_2017}.
In LBM, the solution of \cref{new_filtered} is split into two parts: a collision step and a streaming step. The proposed model modifies only the collision step, which becomes
\begin{equation}\label{collision_model}
    \overline {f_i}^{\rm\, post} =\overline {{f}_i}-\,\underbrace{\omega \,\overline{f^\one_i}}_{\text{\cref{f1approx1}}}-\underbrace{\,\omega_t \, f_{\sgs,i}}_{\text{\cref{turbulent_viscosity,closure1}}}
\end{equation}
Here, \(\overline{f_i}^{\rm\, post}\) is the post-collision value of the discrete distribution function \(\overline {{f}_i}\), while \(\overline{f^\one_i}\) and \( f_{\sgs,i}\) denote the discrete counterparts of \(\overline{f^\one}\) and \( f_{\sgs}^\zero\), computed using the discrete dimensional forms of \cref{f1approx1,closure1}, i.e., in lattice units and $\epsilon = 1$ (see e.g. \cite{kruger_lattice_2017} for discretization details), and $\omega_t$ is computed from \cref{turbulent_viscosity}.
In the code, $H_{i\alpha_1\alpha_2}$ was replaced by $\xi_{i\alpha_1}\xi_{i\alpha_2}$ without the extra trace that vanishes in the incompressible limit.

We tested the model using definition (b) in \cref{turbulent_viscosity}, although (a) is also viable in principle. In the reported cases, the scheme remains stable and gives the expected trends, provided that \(\nu_t\) is prevented from becoming too small or negative.
We adopt a purely dissipative configuration by imposing \(\nu_t \geq \nu\). This condition effectively acts as a criterion for activating the turbulent collision model.
In general, allowing \(\nu_t < \nu\) at coarse resolutions (large filter lengths) promotes turbulence development and mitigates excessive filtering. However, this introduces convergence issues at near-resolved resolutions, because \(f_\sgs\) does not vanish. Physically, \cref{f1approx1,closure1} injects into \(f_\sgs\) both hydrodynamic and nonhydrodynamic contributions, which remain nonzero even at fully resolved scales and may diverge when relaxed through \cref{turbulent_viscosity}. Numerically, \(f_\sgs\) also contains truncation errors that are amplified when \(\nu_t\) is small.
Thus, at coarse resolutions, \(f_\sgs\) primarily represents turbulent fluctuations, while at fine resolutions it becomes dominated by numerical errors and higher-order terms, which negatively impact stability and accuracy.

\begin{figure}
    \centering
    \includegraphics[trim={0 2mm 0 5mm},clip,scale=0.9]{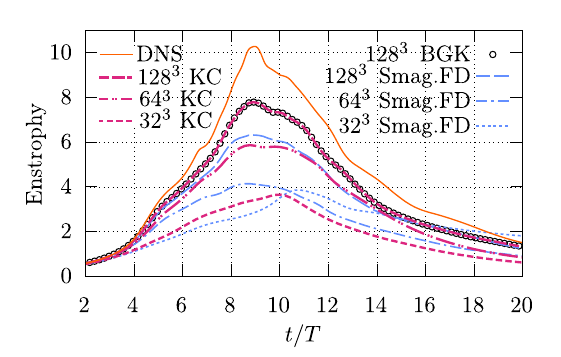}
    \caption{Enstrophy evolution in the TGV test case.}
    \label{tgv1600}
\end{figure}

\enlargethispage{\baselineskip}
The collision model \cref{collision_model} was tested in two canonical turbulent configurations: the Taylor--Green vortex (TGV) \cite{taylor_mechanism_1936,dairay_numerical_2017,laizet_3d_2019} and the turbulent mixing layer (ML) \cite{rogers_direct_1994}. For both test cases, results are shown for: the KC of \cref{collision_model} based on definition (b) in \cref{turbulent_viscosity} imposing \(\nu_t \geq \nu\); for the Smagorinsky model; and for the DNS solution of reference \cite{dairay_numerical_2017,laizet_3d_2019}.  We used the multi-GPU version of the open-source library PALABOS \cite{latt_palabos_2021,latt_multi-gpu_2025}.
For both the Smagorinsky and KC we used a second-order finite differences computation of the velocity gradients and the \(\mathrm{D3Q27}\) lattice.
The KC has been implemented assuming the classical trapezoidal redefinition of the populations appearing in \cref{collision_model} that leads to a redefinition of the viscosity: $\nu_t = c_s^2 (1/\omega_t-1/2)$, with $c_s^2$ being the lattice speed of sound.
In the evaluation of \cref{turbulent_viscosity}, the reported simulations used the reference density \(\rho_0=1\) in lattice units in place of the local \(\bar\rho\).

For the TGV, we report in \cref{tgv1600} the time evolution of the integral enstrophy
\(
    \mathcal{Z}(t) = \tfrac{1}{2} \int \left(\partial_i u_j \, \partial_i u_j - \partial_i u_j \, \partial_j u_i \right) \, \dd \bm{x},
\) computed with a 6th-order finite difference stencil,
at \(\Re=1600\) and \(\Ma=0.2\).
The model constants, $C_s=0.105$ for the Smagorinsky model and $C_\nu=0.015$ for KC, were chosen as the minimum values, reducing in steps of $\Delta C_{s/\nu}=0.005$, that ensured the stability of the simulation at a $32\times32\times32$ resolution.
The results show that KC is significantly less dissipative than the Smagorinsky model and converges toward the BGK model at higher resolutions.

\begin{figure}
    \centering
    \includegraphics[trim={0 1mm 0 2mm},clip,scale=0.9]{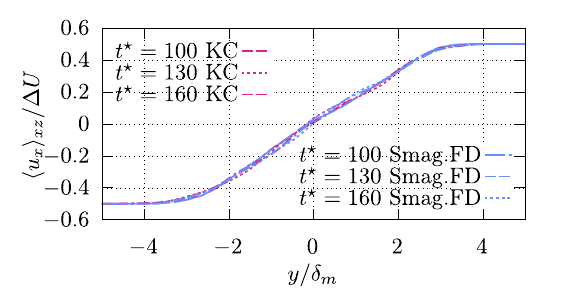}
    \caption{Velocity self-similarity in the ML test case.}
    \label{ml}
\end{figure}

For the turbulent ML, we verified the self-similarity of the velocity profile in the configuration of \cite{malaspinas_consistent_2012}, except that we consider a double mixing layer and impose periodic boundary conditions even in the crossflow direction (\(y\)). The initial velocity field is
\(
    u_x(y) = \tfrac{1}{2} \,\Delta U \, \Bigl[\operatorname{erf}\!\left(
        \frac{y-L_y/4}{\sqrt{2 \pi} \, \delta_0}
        \right) - \operatorname{erf}\!\left(
        \frac{y-3L_y/4}{\sqrt{2 \pi} \, \delta_0}
    \right)\Bigr] - \tfrac{\Delta U}{2},
\)
where \(\delta_0 \equiv \delta_m(t=0)= L_y/100\) is the initial momentum thickness, \(L_y=2L_x=2L_z=128\) is the domain size in lattice units, and \(\Delta U\) is the velocity difference between the two counter-moving streams, here set to \(0.05\) in lattice units. The viscosity is determined by the Reynolds number
\(
    \Re = \sfrac{\Delta U \, \delta_0}{\nu} = 800.
\)
\Cref{ml} shows the normalized velocity profiles, obtained by averaging in the \emph{xz}--plane and scaling with the similarity coordinate defined from the evolving boundary-layer thickness.

To conclude, the FBGK--BE retains information about subfilter scales and accounts for their advection, unlike the filtered NSE, enabling the development of less diffusive numerical models.
Subfilter diffusion, however, still requires modeling: this calls for a generalization of the BGK collision, since the filtered equilibrium cannot be regarded as constant during the collision process.
It is possible to interpret the BGK model based on filtered conserved moments ($\bar \rho, \tilde u, \tilde \theta$) as a naive turbulence closure whose hydrodynamic limit converges to the filtered NSE but that fails to capture the dependence of subfilter dissipation on the subfilter stress tensor.
We showed that this generalization yields an effective KC, which requires only the velocity gradient to separate filtered and subfilter contributions.
A first implementation of the KC shows stable behavior with reduced dissipation compared to the Smagorinsky model. Its underlying principle is general and may be extended to thermal flows and incorporated into more advanced collision operators, which is left as future work.

The authors gratefully acknowledge support from the Swiss National Science Foundation (Grant No. 212882, Advances in turbulence modeling with the lattice Boltzmann method).

\clearpage
\onecolumngrid
\section*{Ethical Use Statement}

The authors ask readers to use the manuscript, methods, numerical models, and results only for peaceful and civil purposes, and not for military or defense purposes, whether directly or indirectly.
\vspace{0.5cm}
\twocolumngrid

\twocolumngrid
\onecolumngrid
\vspace{.3cm}
\noindent\rule{\linewidth}{0.4pt}
\vspace*{-.4cm}
\section*{Backmatter}
\twocolumngrid

\appendix
\setcounter{secnumdepth}{2}
\section{Chapman--Enskog expansion, hydrodynamic limit and kinetic closure of the filtered Boltzmann equation}
\label{CE}

In the CE expansion framework, $\overline f$, and the space and time derivatives are respectively expanded as $\overline f=\sum_{k=0}^{\infty} \epsilon^k \overline{f^{(k)}}$, $\partial_\alpha^{\star}=\sum_{k=1}^{\infty} \frac{{\Ma^2}}{\Re_{\ell}} \epsilon^{k-1} \partial_\alpha^{(k)}$ and $\partial_t^{\star}=\sum_{k=1}^{\infty} \frac{{\Ma^2}}{\Re_{\ell}} \epsilon^{k-1} \partial_t^{(k)}$, with $\epsilon\ll 1$.
The expression of the smallness parameter $\epsilon$ follows directly from the choice of nondimensional numbers used in the scaling, namely \(\epsilon = \Ma^2 / \Re_{\ell}\).
In the literature, this is often informally taken as \(\epsilon \sim \Kn \sim \Ma / \Re\), where \(\Kn\) denotes the Knudsen number.
In the expansion of the derivatives, the prefactor ${{\Ma^2}}/{\Re_{\ell}}$ arises from the term ${\Re_{\ell}}/{{\Ma^2}}$ appearing in front of the collision operator in \cref{fbe_bgk}.
Similarly, the expansion of the collision operator $\overline{\Omega^*}=\sum_{k=0}^{\infty} \epsilon^k \overline{\Omega^{*(k)}}=-\sum_{k=1}^{\infty} \epsilon^k\omega^*\overline{f^{(k)}} -\sum_{k=0}^{\infty} \epsilon^k(\omega_t f_\sgs)^{(k)}$ leads to the following set of equations when the expansions are substituted into \cref{new_filtered}:
\begin{equation}
    0  =(\omega_t f_\sgs)^\zero   \label{fce0}
\end{equation}
\begin{alignat}{2}
     \epsilon\, \partial_t^{(1)} \overline {f^{(0)}}+\xi_{\alpha}^\star \epsilon\, \partial_\alpha^{(1)} \overline {f^{(0)}} +\epsilon\,(\omega_t f_\sgs)^\one& = -\epsilon\,\omega^*\overline{f^{(1)}}\, & &  \label{fce1}\\[.5em]
    \epsilon^2 \partial_t^{(1)} \overline {f^{(1)}}+ \epsilon^2 \partial_t^{(2)} \overline {f^{(0)}}+&\nonumber\\
    \xi_{\alpha}^\star \epsilon^2 \partial_\alpha^{(1)} \overline {f^{(1)}}+\xi_{\alpha}^\star \epsilon^2 \partial_\alpha^{(2)} \overline{f^{(0)}} +&& & \label{fce2}\\
    \epsilon^2\,(\omega_t f_\sgs)^\two  &=-\epsilon^2\,\omega^*\overline{f^{(2)}}\,,\nonumber
\end{alignat}
up to second order in $\epsilon$, the smallness parameter. The expression of \(\overline{f^\one}\) differs from that of the unfiltered case and is given by:
\begin{alignedEq}\label{approx1}
    \epsilon \overline{f^\one} &= \overline f-\underline f^\zero - f_\sgs - \sum_{k=2}^\infty  \epsilon^k \overline{f^{(k)}} \\
    &\approx \overline f-\underline f^\zero - f_\sgs
    \\
    &\approx \underbrace{ \overline f-\underline f^\zero - f_\sgs^\zero}_{\epsilon f^\one} - \epsilon f_\sgs^\one.
\end{alignedEq}
Here, we have defined $f^\one$ such that it is distinct from the computable non-equilibrium component $f^\Neq = \overline f-\underline f^\zero$, which is generally $O(1)$ rather than $O(\epsilon)$.
It is important to emphasize that in \cref{fce1,fce2}, no scale separation between turbulent fluctuations and the mean flow is assumed.
Unlike previous studies~\cite{chen_expanded_2004}, the CE analysis here is applied in the classical sense, separating diffusive from convective dynamics as in the standard BE, rather than splitting turbulent fluctuations from the main flow.

By taking the first-order moment of \cref{fce2} after expressing $\overline{f^\one}$ using \cref{fce1}, we obtain:
\begin{alignatEq}{2}
    \label{filtered_second_moments}
    0 =\ &
    \T[\pp^\two][2][2][] \T[m^{\zero }][1][2][] +
    \pp_t^\two \T[m^{\zero }][1][1][]+\T[\pp^\two][2][2][] \T[m^{\sgs}][1][2][] \\
    &\underbracea{-
        \tau^* \T[\pp^{(1)}][2][3][] \T[m^{\zero}][1][3][]
        -\tau^*\pp_{\alpha_2}^{(1)}
        \pp_t^{(1)}\T[m^\zero ][1][2]}\\
        &
        \pushright{\underbracebd{-\tau^* \T[\pp^{(1)}][2][3][] \T[m^{\sgs}][1][3][]
        -\tau^*\pp_{\alpha_2}^{(1)}\pp_t^{(1)}\T[m^{\sgs}][1][2]}_{\phantom{..}\pp_{\alpha_2}^{(1)}(\bullet)\hfill}}\\
        &-\tau^*\T[\pp^\one][2][2][] \left(\omega_t \T[m^{\sgs}][1][2]\right)^\one\,,
\end{alignatEq}
where we considered $\tau^* = 1/\omega^*$ uniform and constant, and $\T[m^\sgs][1][n]=\int_\Xi {f_\sgs}\xi_{\alpha_1}^\star..\xi_{\alpha_n}^\star\dd\bm\xi^\star$ denotes the $n$-th order raw moment tensor of ${f_\sgs^\zero}$ (recalling that $\T[m^\sgs][1][1]=0$).
The last line in \cref{filtered_second_moments} originates from the third term on the LHS of \cref{fce1}, namely $\overline{\mathcal E}_{\rm BGK}$, and, as will become evident, is crucial for the convergence to the filtered NSE.
The underbraced term $\pp_{\alpha_2}^{(1)}(\bullet)$ can be rewritten in two different ways.

The first way exploits the 0-th and 1st raw moments of \cref{fce1} to simplify the derivatives of the higher-order raw moments in \cref{filtered_second_moments} for an isothermal and incompressible flow.
This procedure follows the spirit of Appendix A.2.2 in \cite{kruger_lattice_2017}, but the presence of subfilter-scale moments introduces an additional level of complexity.
After rewriting $(\bullet)$ and recombining the scales by summing the resulting equation with the first-order moment of \cref{fce1}, one obtains the hydrodynamic limit of the FBE:
\begin{equation}\begin{aligned}\label{HLBE}
    0 =
\T[\pp][2][2][] \left(
    \bar \rho
        \,\MT[\tilde u^\star][1] \MT[\tilde u^\star][2]\right)
        +
        \T[\pp][1][1][]\bar p^*
+
\pp_t \left(\bar \rho   \MT[\tilde u^\star][1]\right)
\\+
\T[\pp][2][2]\left[\bar\rho(\Wtilde{\MT[u^\star][1]\MT[u^\star][2]}-\MT[\tilde u^\star][1]\MT[\tilde u^\star][2])
\right]
\\
- \frac{1}{\Re}\,
    \MT[\pp^{(1)}][2]
\left[\bar \rho
\left(
    \MT[\pp^{(1)}][1]\MT[\tilde u^\star][2]
    +\MT[\pp^{(1)}][2]\MT[\tilde u^\star][1]
\right)
\right]        \\
        +\Big[\MT[\pp^\one][2]\big(\MT[\tilde u^\star][1] \MT[\pp^\one][3]\MT[m^{\sgs}][2][3]\big)
        +\MT[\pp^\one][2]\big(\MT[\tilde u^\star][2]\MT[\pp^\one][3]\MT[m^{\sgs}][1][3]\big)
        \\
        -  \T[\pp^{(1)}][2][3][] \T[m^{\sgs}][1][3][]
        - \pp_{\alpha_2}^{(1)}\pp_t^{(1)}\T[m^{\sgs}][1][2]
        \\
        -\T[\pp^\one][2][2][] \left(\omega_t \T[m^{\sgs}][1][2]\right)^\one\Big]\frac{\Ma^2}{\Re\,\tilde\theta^*}
\,,
\end{aligned}
\end{equation}
where $\Re \equiv \sfrac{\Re_{\ell}}{\tau^*\tilde \theta^*}$ and $\bar p^*\equiv\bar \rho \tilde \theta^* \Ma^{-2}$. Here, lines 3--5 correspond to the term $\pp_{\alpha_2}^{(1)}(\bullet)$ in \cref{filtered_second_moments}, while the last line corresponds to the macroscopic effect of $\overline{\mathcal E}_{\rm BGK}$.

The prefactor $\sfrac{\Ma^2}{\Re\,\tilde\theta^*\tau^*}=\sfrac{\Ma^2}{\Re_{\ell}}$ arises from the recombination of the $O(\epsilon)$ and $O(\epsilon^2)$ equations.
However, its presence alone does not justify neglecting the last three lines of \cref{HLBE} relative to the first three.
In the isothermal case considered here, the thermal contributions cancel from the second- and third-order SGS equilibrium commutation moments.
This cancellation does not establish uniform $O(1)$ bounds on the differentiated SGS moments, particularly the time derivative of the SGS stress, or on the modeled collision-source moment and its divergence.
The Newtonian stress in the third line arises from the $O(\Ma^{-2})$ part of the filtered fine-grained nonequilibrium moment $\overline{\MT[m^{\one}][1][2]}$.
One can clearly see the origin of the $O(\Ma^{-2})$ contribution in the expression of the nondimensional second-order raw moment of the resolved equilibrium,
\begin{equation}
        \MT[m^{\zero}][1][2]  \equiv \int_{\Xi} \xi^\star_{\alpha_1}\xi^\star_{\alpha_2} \underline f^\zero \dd\bm\xi^\star=\bar \rho \left(\MT[\tilde u^\star][1] \MT[\tilde u^\star][2]+\frac{\tilde\theta^* \MT[\delta][1][2]}{\Ma^2}\right). \label{M2}
\end{equation}
The second approach to rewriting $(\bullet)$ proceeds by considering the second-order raw moment of \cref{fce1}:
\begin{alignedEq}
	    \overline{\T[m^\one][1][2]} +\tau^*\left(\omega_t \T[m^{\sgs}][1][2]\right)^\one &=\\
	    &\underbracea{-\tau^*\MT[\pp^{(1)}][3]\MT[m^{\zero}][1][2][3]-
	    \tau^*\pp_{t}^{(1)} \MT[m^{\zero}][1][2]}
	    \\
	    &
	    \underbracebd{-\tau^*\MT[\pp^{(1)}][3]\MT[m^{\sgs}][1][2][3]-
	        \tau^*\MT[\pp^{(1)}][t]\MT[m^{\sgs}][1][2]
    }_{\phantom{..}(\bullet)\hfill}.
\end{alignedEq}
Therefore, by comparison with \cref{filtered_second_moments} and \cref{HLBE} (lines 3--5), we can write:
\begin{equation}\label{approx2}
\begin{aligned}
	    \overline{\T[m^\one][1][2]} +\tau^*\left(\omega_t \T[m^{\sgs}][1][2]\right)^\one = &-\frac{\tau^*{\tilde\theta^*}}{\Ma^2}\left[\bar \rho
	    \left(
	    \MT[\pp^{(1)}][1]\MT[\tilde u^\star][2]
	    +\MT[\pp^{(1)}][2]\MT[\tilde u^\star][1]
	    \right)
	    \right]\\&+\tau^*\MT[\tilde u^\star][1] \MT[\pp^\one][3]\MT[m^{\sgs}][2][3]
	        \\&+\tau^*\MT[\tilde u^\star][2]\MT[\pp^\one][3]\MT[m^{\sgs}][1][3]
	        \\
	        &-  \tau^*\T[\pp^{(1)}][3][3][] \T[m^{\sgs}][1][3][]
	        \\&- \tau^*\pp_t^{(1)}\T[m^{\sgs}][1][2].
\end{aligned}
\end{equation}
We now address the specific form and scaling of \(\omega_t \T[m^{\sgs}][1][2]\) by decomposing it as:
\begin{equation}\label{Edecomposition}
    \left(\omega_t \T[m^{\sgs}][1][2]\right)^\one = \omega_t^\zero \T[m^{\sgs\one}][1][2] + \omega_t^\one \T[m^{\sgs\zero}][1][2]\,.
\end{equation}
The ambiguity in \cref{Edecomposition} arises from the dependence of $f_\sgs$ on the flow conditions and filter length, as well as the functional dependence of $\omega_t$ on $f_\sgs$.
To clarify and simplify the analysis, we distinguish two limiting regimes:
\begin{enumerate}
    \item $\T[m^{\sgs\zero}][1][2] = 0,\ \omega_t^\zero\neq 0$: In this regime, subgrid turbulence fluctuations are negligible. Consequently, there is no substantial subgrid turbulent transport to unresolved scales, i.e.:
    \begin{equation}
        \T[\pp][2][2]\left[\bar\rho(\Wtilde{\MT[u^\star][1]\MT[u^\star][2]}-\MT[\tilde u^\star][1]\MT[\tilde u^\star][2])
\right] \approx 0\,.
    \end{equation}
    \item $\T[m^{\sgs\zero}][1][2] \neq 0$: This represents the regime of primary interest, characterized by a substantial amount of unresolved turbulent fluctuations. However, in this regime, we must have $\omega_t^\zero \approx 0$ as a consequence of \cref{fce0}.
\end{enumerate}
We can now decompose and simplify the LHS of \cref{approx2} as follows:
\begin{gather}
      \overline{\T[m^\one][1][2]} +\tau^*\left(\omega_t \T[m^{\sgs}][1][2]\right)^\one
     \approx \overline{\T[m^\one][1][2]} +\tau^*\omega_t^\one \T[m^{\sgs\zero}][1][2]\\
      \overline{\T[m^\one][1][2]}\approx{\T[m^\one][1][2]}-{\T[m^{\sgs\one}][1][2]}\,,
\end{gather}
where $\T[m^\one][1][2]=\int_\Xi {f^\one}\xi_{\alpha_1}^\star\xi_{\alpha_2}^\star\dd\bm\xi^\star$ with ${f^\one}$ defined in \cref{approx1} and $\omega^*{\T[m^\one][1][2]}$ cannot be further decomposed into ${\T[m^\Neq][1][2]}-{\T[m^{\sgs\zero}][1][2]}$ here because these two components, taken alone, are $O(1)$.

\paragraph{Convergence to the filtered NSE}
From this point, one can attempt to disentangle the terms in \cref{approx2}.
A reasonable separation of \cref{approx2} is the following:
\begin{align}
    \overbrace{\omega^*{\T[m^\one][1][2]}- \omega^*{\T[m^{\sgs\one}][1][2]}}^{\omega^*\overline{\T[m^\one][1][2]}}\approx &-\frac{\tilde\theta^*}{\Ma^2}\left[\bar \rho
    \left(
    \MT[\pp^{(1)}][1]\MT[\tilde u^\star][2]
    +\MT[\pp^{(1)}][2]\MT[\tilde u^\star][1]
    \right)
    \right]\label{aneq}
    \\
     {\omega_t^\one}\MT[m^{\sgs\zero}][1][2] \approx &+\MT[\tilde u^\star][1] \MT[\pp^\one][3]\MT[m^{\sgs}][2][3] \nonumber
        \\&+\MT[\tilde u^\star][2]\MT[\pp^\one][3]\MT[m^{\sgs}][1][3]
        \label{msgs} \\
        &-  \T[\pp^{(1)}][3][3][] \T[m^{\sgs}][1][3][]
        - \pp_t^{(1)}\T[m^{\sgs}][1][2],\nonumber
\end{align}
which connects the relaxation process characterized by the collisional relaxation frequency toward $\overline{f^\zero}$ with the macroscopic stress tensor.
Here $\omega_t^\one$ functions as the constitutive parameter governing the irreversible dissipation; consequently, if one assumes the disentanglement of \cref{aneq,msgs} to be true, a vanishing frequency ($\omega_t^\one \to 0$) implies a collisionless, reversible limit for the subgrid transport terms that precludes the necessary thermalization of cascading energy, inevitably resulting in unphysical spectral accumulation in the higher spectral region.

The separation of \cref{approx2} into \cref{aneq,msgs} is not unique. The present split is one sufficient decomposition for the hydrodynamic limit of the FBE to recover the filtered NSE. It becomes necessary only once the constitutive gauge is fixed by requiring that no additional divergence-free stress term be reassigned between the two parts of the split. In fact, if we assume \cref{msgs} holds and inject it into \cref{HLBE}, we obtain:
\begin{equation}\label{FNSE}
\begin{aligned}
    0 =
\T[\pp][2][2][] \left(
    \bar \rho
        \,\MT[\tilde u^\star][1] \MT[\tilde u^\star][2]\right)
        +
        \T[\pp][1][1][]\bar p^*
+
\pp_t \left(\bar \rho   \MT[\tilde u^\star][1]\right)
\\+
\T[\pp][2][2]\left[\bar\rho(\Wtilde{\MT[u^\star][1]\MT[u^\star][2]}-\MT[\tilde u^\star][1]\MT[\tilde u^\star][2])
\right]
\\
- \frac{1}{\Re}\,
    \MT[\pp^{(1)}][2]
\left[\bar \rho
\left(
    \MT[\pp^{(1)}][1]\MT[\tilde u^\star][2]
    +\MT[\pp^{(1)}][2]\MT[\tilde u^\star][1]
\right)
\right]\mathrlap{,}
\end{aligned}
\end{equation}
which is exactly the (isothermal, incompressible) filtered NSE.

Within the chosen disentanglement, \cref{aneq,msgs} constitute the central result of this work, as they provide a sufficient split for separating filter-scale from subfilter-scale effects.
Formally, under this split, the Newtonian constitutive law for the resolved stress holds even in the filtered case.
It is important to emphasize that \cref{msgs} does not define \(\omega_t\); rather, it demonstrates that, at the macroscopic level, the distinct relaxation time associated with \(f_\sgs^\zero\) constrains the evolution of \(\MT[m^\sgs][1][2]\).
As a consequence, \(\omega_t\) must still be modeled through an appropriate phenomenological turbulence closure.

Beyond this, \cref{aneq,msgs} also provide the foundation for constructing the kinetic turbulence closure.
In particular, they enable a first-order approximation of \(f^\one\) via a Hermite expansion in the multivariate form of Grad~\cite{hermite1864series,grad_note_1949,shan_kinetic_2006}:
\begin{equation}\label{f1}
    \overline{f^\one}
    = w \sum_{n=2}^{\infty} \frac{\Ma^n}{n! (\theta_R^*)^n}\,\T[H][1][n]\,
      \overline{\T[a^\one][1][n]},
\end{equation}
where \(w \equiv w^\star=\Ma^3(2\pi\theta_R^*)^{-3/2}\exp\!\left( - \sfrac{\xi^{*2}}{2\theta_R^*} \right)\) is the fixed-reference Hermite weight written as a convective velocity-space density, with \(\xi_\alpha^*\equiv \xi_\alpha/\sqrt{\theta_R}\),
\(\T[H][1][n]\) are the multivariate Hermite polynomials in the diffusive variable \(\xi^*\), with \(H_{\alpha_1\alpha_2}=\xi_{\alpha_1}^*\xi_{\alpha_2}^*-\theta_R^*\delta_{\alpha_1\alpha_2}\), while the second-order coefficient used below is written in convective units, \(\overline{a^\one_{\alpha_1\alpha_2}}=\overline{m^\one_{\alpha_1\alpha_2}}\equiv\int_\Xi \overline{f^\one}\xi^\star_{\alpha_1}\xi^\star_{\alpha_2}\dd\bm\xi^\star\) for incompressible flows (see \cite{shan_kinetic_2006,malaspinas_lattice_2009,malaspinas_increasing_2015} for details).
Here \(\theta_R^*=1\) in the present nondimensionalization, but it is retained as a placeholder to keep the fixed-reference Hermite scaling explicit.
The factor \(\Ma^n\) appears because the Hermite coefficients are written in convective units.
Truncating \cref{f1} at second order and applying \cref{aneq} directly yields \cref{f1approx1} and, consequently, \cref{closure1}.

If recursive regularization formulas~\cite{shan_kinetic_2006,malaspinas_increasing_2015} are verified to hold in the filtered case, higher-order approximations could be systematically obtained. Furthermore, the present discussion naturally extends to a multi-relaxation-time collision matrix framework.

\paragraph{Alternative disentanglement}
Even after excluding divergence-free stress reassignments, \cref{aneq,msgs} are not unique unless one imposes the filtered NSE, and thus the Newtonian constitutive law, from the outset. For example, one could alternatively consider the following ansatz:
\begin{equation}
    \omega^*{\T[m^\one][1][2]} \approx -\frac{\tilde\theta^*}{\Ma^2}\left[\bar \rho
    \left(
    \MT[\pp^{(1)}][1]\MT[\tilde u^\star][2]
    +\MT[\pp^{(1)}][2]\MT[\tilde u^\star][1]
    \right)
    \right]\label{aneq2}
\end{equation}
    \begin{align}
     \omega_t^\one \T[m^{\sgs\zero}][1][2]-\omega^*{\T[m^{\sgs\one}][1][2]} \approx &+\MT[\tilde u^\star][1] \MT[\pp^\one][3]\MT[m^{\sgs}][2][3] \nonumber
        \\&+\MT[\tilde u^\star][2]\MT[\pp^\one][3]\MT[m^{\sgs}][1][3]
        \label{msgs2} \\
        &-  \T[\pp^{(1)}][3][3][] \T[m^{\sgs}][1][3][]
        - \pp_t^{(1)}\T[m^{\sgs}][1][2].\nonumber
\end{align}
This leads to the macroscopic equation
\begin{equation}\label{FNSE2}
\begin{aligned}
    0 =
\T[\pp][2][2][] \left(
    \bar \rho
        \,\MT[\tilde u^\star][1] \MT[\tilde u^\star][2]\right)
        +
        \T[\pp][1][1][]\bar p^*
+
\pp_t \left(\bar \rho   \MT[\tilde u^\star][1]\right)
\\+
\T[\pp][2][2]\left[\bar\rho(\Wtilde{\MT[u^\star][1]\MT[u^\star][2]}-\MT[\tilde u^\star][1]\MT[\tilde u^\star][2])
\right]
\\
- \frac{1}{\Re}\,
    \MT[\pp^{(1)}][2]
\left[\bar \rho
\left(
    \MT[\pp^{(1)}][1]\MT[\tilde u^\star][2]
    +\MT[\pp^{(1)}][2]\MT[\tilde u^\star][1]
\right)
\right]\\
-\frac{\Ma^2\omega^*}{\Re\,\tilde\theta^*}\MT[\pp^{(1)}][2]{\T[m^{\sgs\one}][1][2]} \,,
\end{aligned}
\end{equation}
where an explicit subgrid dissipation term appears. While the full investigation of this alternative hydrodynamic limit is left for future work, two key observations emerge: (i) operationally, this formulation shares the same closure challenges as \cref{aneq,msgs,FNSE}, as both require estimating $f^\one \neq f^\Neq$ and $\omega_t$; but (ii) theoretically, \cref{msgs2} yields subgrid dissipation even when $\omega_t^\one = 0$. This invalidates the assumption that $\omega_t \sim O(\epsilon)$ is strictly necessary for thermodynamic consistency. However, since numerical experiments indicate that $\omega_t = 0$ leads to unstable simulations, this scaling remains a practical requirement for numerical stability.


\begin{thebibliography}{53}%
\makeatletter
\providecommand \@ifxundefined [1]{%
 \@ifx{#1\undefined}
}%
\providecommand \@ifnum [1]{%
 \ifnum #1\expandafter \@firstoftwo
 \else \expandafter \@secondoftwo
 \fi
}%
\providecommand \@ifx [1]{%
 \ifx #1\expandafter \@firstoftwo
 \else \expandafter \@secondoftwo
 \fi
}%
\providecommand \natexlab [1]{#1}%
\providecommand \enquote  [1]{``#1''}%
\providecommand \bibnamefont  [1]{#1}%
\providecommand \bibfnamefont [1]{#1}%
\providecommand \citenamefont [1]{#1}%
\providecommand \href@noop [0]{\@secondoftwo}%
\providecommand \href [0]{\begingroup \@sanitize@url \@href}%
\providecommand \@href[1]{\@@startlink{#1}\@@href}%
\providecommand \@@href[1]{\endgroup#1\@@endlink}%
\providecommand \@sanitize@url [0]{\catcode `\\12\catcode `\$12\catcode
  `\&12\catcode `\#12\catcode `\^12\catcode `\_12\catcode `\%12\relax}%
\providecommand \@@startlink[1]{}%
\providecommand \@@endlink[0]{}%
\providecommand \url  [0]{\begingroup\@sanitize@url \@url }%
\providecommand \@url [1]{\endgroup\@href {#1}{\urlprefix }}%
\providecommand \urlprefix  [0]{URL }%
\providecommand \Eprint [0]{\href }%
\providecommand \doibase [0]{https://doi.org/}%
\providecommand \selectlanguage [0]{\@gobble}%
\providecommand \bibinfo  [0]{\@secondoftwo}%
\providecommand \bibfield  [0]{\@secondoftwo}%
\providecommand \translation [1]{[#1]}%
\providecommand \BibitemOpen [0]{}%
\providecommand \bibitemStop [0]{}%
\providecommand \bibitemNoStop [0]{.\EOS\space}%
\providecommand \EOS [0]{\spacefactor3000\relax}%
\providecommand \BibitemShut  [1]{\csname bibitem#1\endcsname}%
\let\auto@bib@innerbib\@empty
\bibitem [{\citenamefont {Pope}(2000)}]{pope_turbulent_2000}%
  \BibitemOpen
  \bibfield  {author} {\bibinfo {author} {\bibfnamefont {S.~B.}\ \bibnamefont
  {Pope}},\ }\href@noop {} {\emph {\bibinfo {title} {Turbulent flows}}}\
  (\bibinfo  {publisher} {Cambridge University Press},\ \bibinfo {address}
  {Cambridge; New York},\ \bibinfo {year} {2000})\BibitemShut {NoStop}%
\bibitem [{\citenamefont {Davidson}(2015)}]{davidson_turbulence_2015}%
  \BibitemOpen
  \bibfield  {author} {\bibinfo {author} {\bibfnamefont {P.}~\bibnamefont
  {Davidson}},\ }\href@noop {} {\emph {\bibinfo {title} {Turbulence: An
  Introduction for Scientists and Engineers}}},\ \bibinfo {edition} {2nd}\ ed.\
  (\bibinfo  {publisher} {Oxford University Press},\ \bibinfo {address}
  {Oxford, United Kingdom; New York, NY, United States of America},\ \bibinfo
  {year} {2015})\BibitemShut {NoStop}%
\bibitem [{\citenamefont {Reynolds}(1895)}]{reynolds_iv_1895}%
  \BibitemOpen
  \bibfield  {author} {\bibinfo {author} {\bibfnamefont {O.}~\bibnamefont
  {Reynolds}},\ }\href {https://doi.org/10.1098/rsta.1895.0004} {\bibfield
  {journal} {\bibinfo  {journal} {Philosophical Transactions of the Royal
  Society of London. (A.)}\ }\textbf {\bibinfo {volume} {186}},\ \bibinfo
  {pages} {123} (\bibinfo {year} {1895})}\BibitemShut {NoStop}%
\bibitem [{\citenamefont {Smagorinsky}(1963)}]{smagorinsky_general_1963}%
  \BibitemOpen
  \bibfield  {author} {\bibinfo {author} {\bibfnamefont {J.}~\bibnamefont
  {Smagorinsky}},\ }\href
  {https://doi.org/10.1175/1520-0493(1963)091<0099:GCEWTP>2.3.CO;2} {\bibfield
  {journal} {\bibinfo  {journal} {Monthly weather review}\ }\textbf {\bibinfo
  {volume} {91}},\ \bibinfo {pages} {99} (\bibinfo {year} {1963})}\BibitemShut
  {NoStop}%
\bibitem [{\citenamefont {Lilly}(1966)}]{lilly_representation_1966}%
  \BibitemOpen
  \bibfield  {author} {\bibinfo {author} {\bibfnamefont {D.}~\bibnamefont
  {Lilly}},\ }\href {https://doi.org/10.5065/D62R3PMM} {\emph {\bibinfo {title}
  {The representation of small-scale turbulence in numerical simulation
  experiments}}},\ \bibinfo {type} {Tech. Rep.}\ (\bibinfo {year}
  {1966})\BibitemShut {NoStop}%
\bibitem [{\citenamefont {Sagaut}(2006)}]{sagaut_large_2006}%
  \BibitemOpen
  \bibfield  {author} {\bibinfo {author} {\bibfnamefont {P.}~\bibnamefont
  {Sagaut}},\ }\href@noop {} {\emph {\bibinfo {title} {Large eddy simulation
  for incompressible flows: an introduction}}},\ \bibinfo {edition} {3rd}\
  ed.,\ Scientific computation\ (\bibinfo  {publisher} {Springer},\ \bibinfo
  {address} {Berlin; New York},\ \bibinfo {year} {2006})\BibitemShut {NoStop}%
\bibitem [{\citenamefont {Kolmogorov}(1941)}]{kolmogorov_local_1941}%
  \BibitemOpen
  \bibfield  {author} {\bibinfo {author} {\bibfnamefont {A.~N.}\ \bibnamefont
  {Kolmogorov}},\ }\href@noop {} {\bibfield  {journal} {\bibinfo  {journal}
  {Doklady Akademii Nauk SSSR}\ }\textbf {\bibinfo {volume} {30}},\ \bibinfo
  {pages} {299} (\bibinfo {year} {1941})}\BibitemShut {NoStop}%
\bibitem [{\citenamefont {Navier}(1822)}]{navier_memoire_1822}%
  \BibitemOpen
  \bibfield  {author} {\bibinfo {author} {\bibfnamefont {C.-L. M.~H.}\
  \bibnamefont {Navier}},\ }\href@noop {} {\bibfield  {journal} {\bibinfo
  {journal} {Mémoires de l’Académie Royale des Sciences de l’Institut de
  France}\ }\textbf {\bibinfo {volume} {6}},\ \bibinfo {pages} {389} (\bibinfo
  {year} {1822})}\BibitemShut {NoStop}%
\bibitem [{\citenamefont {Stokes}(1845)}]{stokes_theories_1845}%
  \BibitemOpen
  \bibfield  {author} {\bibinfo {author} {\bibfnamefont {G.~G.}\ \bibnamefont
  {Stokes}},\ }\href@noop {} {\bibfield  {journal} {\bibinfo  {journal}
  {Transactions of the Cambridge Philosophical Society}\ }\textbf {\bibinfo
  {volume} {8}},\ \bibinfo {pages} {287} (\bibinfo {year} {1845})}\BibitemShut
  {NoStop}%
\bibitem [{\citenamefont {Rotta}(1951)}]{rotta_statistische_1951}%
  \BibitemOpen
  \bibfield  {author} {\bibinfo {author} {\bibfnamefont {J.}~\bibnamefont
  {Rotta}},\ }\href {https://doi.org/10.1007/BF01330059} {\bibfield  {journal}
  {\bibinfo  {journal} {Zeitschrift für Physik}\ }\textbf {\bibinfo {volume}
  {129}},\ \bibinfo {pages} {547} (\bibinfo {year} {1951})}\BibitemShut
  {NoStop}%
\bibitem [{\citenamefont {Ferziger}\ \emph {et~al.}(2020)\citenamefont
  {Ferziger}, \citenamefont {Perić},\ and\ \citenamefont
  {Street}}]{ferziger2020computational}%
  \BibitemOpen
  \bibfield  {author} {\bibinfo {author} {\bibfnamefont {J.~H.}\ \bibnamefont
  {Ferziger}}, \bibinfo {author} {\bibfnamefont {M.}~\bibnamefont {Perić}},\
  and\ \bibinfo {author} {\bibfnamefont {R.~L.}\ \bibnamefont {Street}},\
  }\href {https://doi.org/10.1007/978-3-319-99693-6} {\emph {\bibinfo {title}
  {Computational Methods for Fluid Dynamics}}},\ \bibinfo {edition} {4th}\ ed.\
  (\bibinfo  {publisher} {Springer International Publishing},\ \bibinfo
  {address} {Cham},\ \bibinfo {year} {2020})\BibitemShut {NoStop}%
\bibitem [{\citenamefont {Boltzmann}(1872)}]{boltzmann_weirere_1872}%
  \BibitemOpen
  \bibfield  {author} {\bibinfo {author} {\bibfnamefont {L.}~\bibnamefont
  {Boltzmann}},\ }\href@noop {} {\bibfield  {journal} {\bibinfo  {journal} {K.
  Acad. Wiss.(Wein) Sitzb., II Abt}\ }\textbf {\bibinfo {volume} {66}}
  (\bibinfo {year} {1872})}\BibitemShut {NoStop}%
\bibitem [{\citenamefont {Bhatnagar}\ \emph {et~al.}(1954)\citenamefont
  {Bhatnagar}, \citenamefont {Gross},\ and\ \citenamefont
  {Krook}}]{bhatnagar_model_1954}%
  \BibitemOpen
  \bibfield  {author} {\bibinfo {author} {\bibfnamefont {P.~L.}\ \bibnamefont
  {Bhatnagar}}, \bibinfo {author} {\bibfnamefont {E.~P.}\ \bibnamefont
  {Gross}},\ and\ \bibinfo {author} {\bibfnamefont {M.}~\bibnamefont {Krook}},\
  }\href {https://doi.org/10.1103/PhysRev.94.511} {\bibfield  {journal}
  {\bibinfo  {journal} {Physical Review}\ }\textbf {\bibinfo {volume} {94}},\
  \bibinfo {pages} {511} (\bibinfo {year} {1954})}\BibitemShut {NoStop}%
\bibitem [{\citenamefont {Ansumali}\ \emph {et~al.}(2004)\citenamefont
  {Ansumali}, \citenamefont {Karlin},\ and\ \citenamefont
  {Succi}}]{ansumali_kinetic_2004}%
  \BibitemOpen
  \bibfield  {author} {\bibinfo {author} {\bibfnamefont {S.}~\bibnamefont
  {Ansumali}}, \bibinfo {author} {\bibfnamefont {I.~V.}\ \bibnamefont
  {Karlin}},\ and\ \bibinfo {author} {\bibfnamefont {S.}~\bibnamefont
  {Succi}},\ }\href {https://doi.org/10.1016/j.physa.2004.02.013} {\bibfield
  {journal} {\bibinfo  {journal} {Physica A: Statistical Mechanics and its
  Applications}\ }\textbf {\bibinfo {volume} {338}},\ \bibinfo {pages} {379}
  (\bibinfo {year} {2004})}\BibitemShut {NoStop}%
\bibitem [{\citenamefont {Malaspinas}\ and\ \citenamefont
  {Sagaut}(2012)}]{malaspinas_consistent_2012}%
  \BibitemOpen
  \bibfield  {author} {\bibinfo {author} {\bibfnamefont {O.}~\bibnamefont
  {Malaspinas}}\ and\ \bibinfo {author} {\bibfnamefont {P.}~\bibnamefont
  {Sagaut}},\ }\href {https://doi.org/10.1017/jfm.2012.155} {\bibfield
  {journal} {\bibinfo  {journal} {Journal of Fluid Mechanics}\ }\textbf
  {\bibinfo {volume} {700}},\ \bibinfo {pages} {514} (\bibinfo {year}
  {2012})}\BibitemShut {NoStop}%
\bibitem [{\citenamefont {Chen}\ \emph {et~al.}(1999)\citenamefont {Chen},
  \citenamefont {Succi},\ and\ \citenamefont {Orszag}}]{chen_analysis_1999}%
  \BibitemOpen
  \bibfield  {author} {\bibinfo {author} {\bibfnamefont {H.}~\bibnamefont
  {Chen}}, \bibinfo {author} {\bibfnamefont {S.}~\bibnamefont {Succi}},\ and\
  \bibinfo {author} {\bibfnamefont {S.}~\bibnamefont {Orszag}},\ }\href
  {https://doi.org/10.1103/PhysRevE.59.R2527} {\bibfield  {journal} {\bibinfo
  {journal} {Physical Review E}\ }\textbf {\bibinfo {volume} {59}},\ \bibinfo
  {pages} {R2527} (\bibinfo {year} {1999})},\ \bibinfo {note} {publisher:
  American Physical Society}\BibitemShut {NoStop}%
\bibitem [{\citenamefont {Succi}\ \emph {et~al.}(2002)\citenamefont {Succi},
  \citenamefont {Filippova}, \citenamefont {Chen},\ and\ \citenamefont
  {Orszag}}]{succi_towards_2002}%
  \BibitemOpen
  \bibfield  {author} {\bibinfo {author} {\bibfnamefont {S.}~\bibnamefont
  {Succi}}, \bibinfo {author} {\bibfnamefont {O.}~\bibnamefont {Filippova}},
  \bibinfo {author} {\bibfnamefont {H.}~\bibnamefont {Chen}},\ and\ \bibinfo
  {author} {\bibfnamefont {S.}~\bibnamefont {Orszag}},\ }\href
  {https://doi.org/10.1023/A:1014570923357} {\bibfield  {journal} {\bibinfo
  {journal} {Journal of Statistical Physics}\ }\textbf {\bibinfo {volume}
  {107}},\ \bibinfo {pages} {261} (\bibinfo {year} {2002})}\BibitemShut
  {NoStop}%
\bibitem [{\citenamefont {Chen}\ \emph {et~al.}(2003)\citenamefont {Chen},
  \citenamefont {Kandasamy}, \citenamefont {Orszag}, \citenamefont {Shock},
  \citenamefont {Succi},\ and\ \citenamefont {Yakhot}}]{chen_extended_2003}%
  \BibitemOpen
  \bibfield  {author} {\bibinfo {author} {\bibfnamefont {H.}~\bibnamefont
  {Chen}}, \bibinfo {author} {\bibfnamefont {S.}~\bibnamefont {Kandasamy}},
  \bibinfo {author} {\bibfnamefont {S.}~\bibnamefont {Orszag}}, \bibinfo
  {author} {\bibfnamefont {R.}~\bibnamefont {Shock}}, \bibinfo {author}
  {\bibfnamefont {S.}~\bibnamefont {Succi}},\ and\ \bibinfo {author}
  {\bibfnamefont {V.}~\bibnamefont {Yakhot}},\ }\href
  {https://doi.org/10.1126/science.1085048} {\bibfield  {journal} {\bibinfo
  {journal} {Science (New York, N.Y.)}\ }\textbf {\bibinfo {volume} {301}},\
  \bibinfo {pages} {633} (\bibinfo {year} {2003})}\BibitemShut {NoStop}%
\bibitem [{\citenamefont {Girimaji}(2007)}]{girimaji_boltzmann_2007}%
  \BibitemOpen
  \bibfield  {author} {\bibinfo {author} {\bibfnamefont {S.~S.}\ \bibnamefont
  {Girimaji}},\ }\href {https://doi.org/10.1103/PhysRevLett.99.034501}
  {\bibfield  {journal} {\bibinfo  {journal} {Physical Review Letters}\
  }\textbf {\bibinfo {volume} {99}},\ \bibinfo {pages} {034501} (\bibinfo
  {year} {2007})}\BibitemShut {NoStop}%
\bibitem [{\citenamefont {Banda}\ \emph {et~al.}(2006)\citenamefont {Banda},
  \citenamefont {Seaïd},\ and\ \citenamefont
  {Teleaga}}]{banda_discrete-velocity_2006}%
  \BibitemOpen
  \bibfield  {author} {\bibinfo {author} {\bibfnamefont {M.}~\bibnamefont
  {Banda}}, \bibinfo {author} {\bibfnamefont {M.}~\bibnamefont {Seaïd}},\ and\
  \bibinfo {author} {\bibfnamefont {I.}~\bibnamefont {Teleaga}},\ }\href
  {https://doi.org/10.1016/j.amc.2006.04.033} {\bibfield  {journal} {\bibinfo
  {journal} {Applied Mathematics and Computation}\ }\textbf {\bibinfo {volume}
  {182}},\ \bibinfo {pages} {739} (\bibinfo {year} {2006})}\BibitemShut
  {NoStop}%
\bibitem [{\citenamefont {Sagaut}(2010)}]{sagaut_toward_2010}%
  \BibitemOpen
  \bibfield  {author} {\bibinfo {author} {\bibfnamefont {P.}~\bibnamefont
  {Sagaut}},\ }\href {https://doi.org/10.1016/j.camwa.2009.08.051} {\bibfield
  {journal} {\bibinfo  {journal} {Computers \& Mathematics with Applications}\
  }\bibinfo {series} {Mesoscopic {Methods} in {Engineering} and {Science}},\
  \textbf {\bibinfo {volume} {59}},\ \bibinfo {pages} {2194} (\bibinfo {year}
  {2010})}\BibitemShut {NoStop}%
\bibitem [{\citenamefont {Malaspinas}\ and\ \citenamefont
  {Sagaut}(2011)}]{malaspinas_advanced_2011}%
  \BibitemOpen
  \bibfield  {author} {\bibinfo {author} {\bibfnamefont {O.}~\bibnamefont
  {Malaspinas}}\ and\ \bibinfo {author} {\bibfnamefont {P.}~\bibnamefont
  {Sagaut}},\ }\href {https://doi.org/10.1063/1.3650422} {\bibfield  {journal}
  {\bibinfo  {journal} {Physics of Fluids}\ }\textbf {\bibinfo {volume} {23}},\
  \bibinfo {pages} {105103} (\bibinfo {year} {2011})}\BibitemShut {NoStop}%
\bibitem [{\citenamefont {Luan}\ \emph {et~al.}(2025)\citenamefont {Luan},
  \citenamefont {Zhang},\ and\ \citenamefont {Zhang}}]{luan_constructing_2025}%
  \BibitemOpen
  \bibfield  {author} {\bibinfo {author} {\bibfnamefont {P.}~\bibnamefont
  {Luan}}, \bibinfo {author} {\bibfnamefont {H.}~\bibnamefont {Zhang}},\ and\
  \bibinfo {author} {\bibfnamefont {J.}~\bibnamefont {Zhang}},\ }\href
  {https://doi.org/10.1017/jfm.2025.323} {\bibfield  {journal} {\bibinfo
  {journal} {Journal of Fluid Mechanics}\ }\textbf {\bibinfo {volume} {1011}},\
  \bibinfo {pages} {A44} (\bibinfo {year} {2025})}\BibitemShut {NoStop}%
\bibitem [{\citenamefont {Chen}\ \emph {et~al.}(2004)\citenamefont {Chen},
  \citenamefont {Orszag}, \citenamefont {Staroselsky},\ and\ \citenamefont
  {Succi}}]{chen_expanded_2004}%
  \BibitemOpen
  \bibfield  {author} {\bibinfo {author} {\bibfnamefont {H.}~\bibnamefont
  {Chen}}, \bibinfo {author} {\bibfnamefont {S.~A.}\ \bibnamefont {Orszag}},
  \bibinfo {author} {\bibfnamefont {I.}~\bibnamefont {Staroselsky}},\ and\
  \bibinfo {author} {\bibfnamefont {S.}~\bibnamefont {Succi}},\ }\href
  {https://doi.org/10.1017/S0022112004001211} {\bibfield  {journal} {\bibinfo
  {journal} {Journal of Fluid Mechanics}\ }\textbf {\bibinfo {volume} {519}},\
  \bibinfo {pages} {301} (\bibinfo {year} {2004})}\BibitemShut {NoStop}%
\bibitem [{\citenamefont {Chen}\ \emph {et~al.}(2013)\citenamefont {Chen},
  \citenamefont {Staroselsky},\ and\ \citenamefont
  {Yakhot}}]{chen_non-perturbative_2013}%
  \BibitemOpen
  \bibfield  {author} {\bibinfo {author} {\bibfnamefont {H.}~\bibnamefont
  {Chen}}, \bibinfo {author} {\bibfnamefont {I.}~\bibnamefont {Staroselsky}},\
  and\ \bibinfo {author} {\bibfnamefont {V.}~\bibnamefont {Yakhot}},\ }\href
  {https://doi.org/10.1088/0031-8949/2013/T155/014040} {\bibfield  {journal}
  {\bibinfo  {journal} {Physica Scripta}\ }\textbf {\bibinfo {volume} {2013}},\
  \bibinfo {pages} {014040} (\bibinfo {year} {2013})}\BibitemShut {NoStop}%
\bibitem [{\citenamefont {Chen}\ \emph {et~al.}(2023)\citenamefont {Chen},
  \citenamefont {Staroselsky}, \citenamefont {Sreenivasan},\ and\ \citenamefont
  {Yakhot}}]{chen_average_2023}%
  \BibitemOpen
  \bibfield  {author} {\bibinfo {author} {\bibfnamefont {H.}~\bibnamefont
  {Chen}}, \bibinfo {author} {\bibfnamefont {I.}~\bibnamefont {Staroselsky}},
  \bibinfo {author} {\bibfnamefont {K.~R.}\ \bibnamefont {Sreenivasan}},\ and\
  \bibinfo {author} {\bibfnamefont {V.}~\bibnamefont {Yakhot}},\ }\href
  {https://doi.org/10.3390/atmos14071109} {\bibfield  {journal} {\bibinfo
  {journal} {Atmosphere}\ }\textbf {\bibinfo {volume} {14}},\ \bibinfo {pages}
  {1109} (\bibinfo {year} {2023})}\BibitemShut {NoStop}%
\bibitem [{\citenamefont {Saveliev}(2024)}]{saveliev_kinetic_2024}%
  \BibitemOpen
  \bibfield  {author} {\bibinfo {author} {\bibfnamefont {V.~L.}\ \bibnamefont
  {Saveliev}},\ }\href {https://doi.org/10.1063/5.0242731} {\bibfield
  {journal} {\bibinfo  {journal} {Physics of Fluids}\ }\textbf {\bibinfo
  {volume} {36}},\ \bibinfo {pages} {125175} (\bibinfo {year}
  {2024})}\BibitemShut {NoStop}%
\bibitem [{\citenamefont {Asinari}\ \emph {et~al.}(2016)\citenamefont
  {Asinari}, \citenamefont {Fasano},\ and\ \citenamefont
  {Chiavazzo}}]{asinari_kinetic_2016}%
  \BibitemOpen
  \bibfield  {author} {\bibinfo {author} {\bibfnamefont {P.}~\bibnamefont
  {Asinari}}, \bibinfo {author} {\bibfnamefont {M.}~\bibnamefont {Fasano}},\
  and\ \bibinfo {author} {\bibfnamefont {E.}~\bibnamefont {Chiavazzo}},\ }\href
  {https://doi.org/10.3390/e18040121} {\bibfield  {journal} {\bibinfo
  {journal} {Entropy}\ }\textbf {\bibinfo {volume} {18}},\ \bibinfo {pages}
  {121} (\bibinfo {year} {2016})},\ \bibinfo {note} {publisher:
  Multidisciplinary Digital Publishing Institute}\BibitemShut {NoStop}%
\bibitem [{\citenamefont {Righi}(2016)}]{righi_gas-kinetic_2016}%
  \BibitemOpen
  \bibfield  {author} {\bibinfo {author} {\bibfnamefont {M.}~\bibnamefont
  {Righi}},\ }\href {https://doi.org/10.1007/s10494-015-9677-2} {\bibfield
  {journal} {\bibinfo  {journal} {Flow, Turbulence and Combustion}\ }\textbf
  {\bibinfo {volume} {97}},\ \bibinfo {pages} {121} (\bibinfo {year}
  {2016})}\BibitemShut {NoStop}%
\bibitem [{\citenamefont {Zhigulev}(1965)}]{zhigulev_equations_1965}%
  \BibitemOpen
  \bibfield  {author} {\bibinfo {author} {\bibfnamefont {V.~N.}\ \bibnamefont
  {Zhigulev}},\ }\href@noop {} {\bibfield  {journal} {\bibinfo  {journal}
  {Doklady Akademii Nauk SSSR}\ }\textbf {\bibinfo {volume} {165}},\ \bibinfo
  {pages} {502} (\bibinfo {year} {1965})}\BibitemShut {NoStop}%
\bibitem [{\citenamefont {Tsugé}(1974)}]{tsuge_approach_1974}%
  \BibitemOpen
  \bibfield  {author} {\bibinfo {author} {\bibfnamefont {S.}~\bibnamefont
  {Tsugé}},\ }\href {https://doi.org/10.1063/1.1694592} {\bibfield  {journal}
  {\bibinfo  {journal} {The Physics of Fluids}\ }\textbf {\bibinfo {volume}
  {17}},\ \bibinfo {pages} {22} (\bibinfo {year} {1974})}\BibitemShut {NoStop}%
\bibitem [{\citenamefont {Chliamovitch}\ \emph {et~al.}(2015)\citenamefont
  {Chliamovitch}, \citenamefont {Malaspinas},\ and\ \citenamefont
  {Chopard}}]{chliamovitch_truncation_2015}%
  \BibitemOpen
  \bibfield  {author} {\bibinfo {author} {\bibfnamefont {G.}~\bibnamefont
  {Chliamovitch}}, \bibinfo {author} {\bibfnamefont {O.}~\bibnamefont
  {Malaspinas}},\ and\ \bibinfo {author} {\bibfnamefont {B.}~\bibnamefont
  {Chopard}},\ }\href {https://doi.org/10.3390/e17117522} {\bibfield  {journal}
  {\bibinfo  {journal} {Entropy}\ }\textbf {\bibinfo {volume} {17}},\ \bibinfo
  {pages} {7522} (\bibinfo {year} {2015})},\ \bibinfo {note} {publisher:
  Multidisciplinary Digital Publishing Institute}\BibitemShut {NoStop}%
\bibitem [{\citenamefont {Chliamovitch}\ \emph {et~al.}(2017)\citenamefont
  {Chliamovitch}, \citenamefont {Malaspinas},\ and\ \citenamefont
  {Chopard}}]{chliamovitch_kinetic_2017}%
  \BibitemOpen
  \bibfield  {author} {\bibinfo {author} {\bibfnamefont {G.}~\bibnamefont
  {Chliamovitch}}, \bibinfo {author} {\bibfnamefont {O.}~\bibnamefont
  {Malaspinas}},\ and\ \bibinfo {author} {\bibfnamefont {B.}~\bibnamefont
  {Chopard}},\ }\href {https://doi.org/10.3390/e19080381} {\bibfield  {journal}
  {\bibinfo  {journal} {Entropy}\ }\textbf {\bibinfo {volume} {19}},\ \bibinfo
  {pages} {381} (\bibinfo {year} {2017})},\ \bibinfo {note} {publisher:
  Multidisciplinary Digital Publishing Institute}\BibitemShut {NoStop}%
\bibitem [{\citenamefont {Favre}(1965)}]{favre_equations_1965}%
  \BibitemOpen
  \bibfield  {author} {\bibinfo {author} {\bibfnamefont {A.}~\bibnamefont
  {Favre}},\ }\href@noop {} {\emph {\bibinfo {title} {Équations des gaz
  turbulents compressibles}}},\ \bibinfo {type} {Tech. Rep.}\ \bibinfo {number}
  {137}\ (\bibinfo  {institution} {SNECMA},\ \bibinfo {address} {France},\
  \bibinfo {year} {1965})\BibitemShut {NoStop}%
\bibitem [{\citenamefont {Favre}(1983)}]{favre_turbulence_1983}%
  \BibitemOpen
  \bibfield  {author} {\bibinfo {author} {\bibfnamefont {A.}~\bibnamefont
  {Favre}},\ }\href {https://doi.org/10.1063/1.864049} {\bibfield  {journal}
  {\bibinfo  {journal} {The Physics of Fluids}\ }\textbf {\bibinfo {volume}
  {26}},\ \bibinfo {pages} {2851} (\bibinfo {year} {1983})}\BibitemShut
  {NoStop}%
\bibitem [{\citenamefont {Enskog}(1922)}]{enskog_kinetische_1922}%
  \BibitemOpen
  \bibfield  {author} {\bibinfo {author} {\bibfnamefont {D.}~\bibnamefont
  {Enskog}},\ }\href@noop {} {\emph {\bibinfo {title} {Kinetische {Theorie} der
  {Wärmeleitung}: {Reibung} und {Selbst}-diffusion in {Gewissen} verdichteten
  gasen und flüssigkeiten}}}\ (\bibinfo  {publisher} {Almqvist \& Wiksells
  boktryckeri-a.-b.},\ \bibinfo {year} {1922})\BibitemShut {NoStop}%
\bibitem [{\citenamefont {Chapman}\ \emph {et~al.}(1953)\citenamefont
  {Chapman}, \citenamefont {Cowling}, \citenamefont {Burnett},\ and\
  \citenamefont {Cercignani}}]{chapman_mathematical_1953}%
  \BibitemOpen
  \bibfield  {author} {\bibinfo {author} {\bibfnamefont {S.}~\bibnamefont
  {Chapman}}, \bibinfo {author} {\bibfnamefont {T.}~\bibnamefont {Cowling}},
  \bibinfo {author} {\bibfnamefont {D.}~\bibnamefont {Burnett}},\ and\ \bibinfo
  {author} {\bibfnamefont {C.}~\bibnamefont {Cercignani}},\ }\href
  {https://books.google.it/books?id=JcjHpiJPKeIC} {\emph {\bibinfo {title} {The
  {Mathematical} {Theory} of {Non}-uniform {Gases}: {An} {Account} of the
  {Kinetic} {Theory} of {Viscosity}, {Thermal} {Conduction} and {Diffusion} in
  {Gases}}}},\ Cambridge {Mathematical} {Library}\ (\bibinfo  {publisher}
  {Cambridge University Press},\ \bibinfo {year} {1953})\BibitemShut {NoStop}%
\bibitem [{\citenamefont {Launder}\ and\ \citenamefont
  {Spalding}(1974)}]{launder_numerical_1974}%
  \BibitemOpen
  \bibfield  {author} {\bibinfo {author} {\bibfnamefont {B.~E.}\ \bibnamefont
  {Launder}}\ and\ \bibinfo {author} {\bibfnamefont {D.~B.}\ \bibnamefont
  {Spalding}},\ }\href {https://doi.org/10.1016/0045-7825(74)90029-2}
  {\bibfield  {journal} {\bibinfo  {journal} {Computer Methods in Applied
  Mechanics and Engineering}\ }\textbf {\bibinfo {volume} {3}},\ \bibinfo
  {pages} {269} (\bibinfo {year} {1974})}\BibitemShut {NoStop}%
\bibitem [{\citenamefont {Malaspinas}(2015)}]{malaspinas_increasing_2015}%
  \BibitemOpen
  \bibfield  {author} {\bibinfo {author} {\bibfnamefont {O.}~\bibnamefont
  {Malaspinas}},\ }\href
  {https://www.researchgate.net/publication/277334522_Increasing_stability_and_accuracy_of_the_lattice_Boltzmann_scheme_recursivity_and_regularization}
  {\bibfield  {journal} {\bibinfo  {journal} {Arxiv}\ } (\bibinfo {year}
  {2015})}\BibitemShut {NoStop}%
\bibitem [{\citenamefont {Higuera}\ \emph {et~al.}(1989)\citenamefont
  {Higuera}, \citenamefont {Succi},\ and\ \citenamefont
  {Benzi}}]{higuera_lattice_1989}%
  \BibitemOpen
  \bibfield  {author} {\bibinfo {author} {\bibfnamefont {F.~J.}\ \bibnamefont
  {Higuera}}, \bibinfo {author} {\bibfnamefont {S.}~\bibnamefont {Succi}},\
  and\ \bibinfo {author} {\bibfnamefont {R.}~\bibnamefont {Benzi}},\ }\href
  {https://doi.org/10.1209/0295-5075/9/4/008} {\bibfield  {journal} {\bibinfo
  {journal} {Europhysics Letters (EPL)}\ }\textbf {\bibinfo {volume} {9}},\
  \bibinfo {pages} {345} (\bibinfo {year} {1989})}\BibitemShut {NoStop}%
\bibitem [{\citenamefont {McNamara}\ and\ \citenamefont
  {Zanetti}(1988)}]{mcnamara_use_1988}%
  \BibitemOpen
  \bibfield  {author} {\bibinfo {author} {\bibfnamefont {G.~R.}\ \bibnamefont
  {McNamara}}\ and\ \bibinfo {author} {\bibfnamefont {G.}~\bibnamefont
  {Zanetti}},\ }\href {https://doi.org/10.1103/PhysRevLett.61.2332} {\bibfield
  {journal} {\bibinfo  {journal} {Physical Review Letters}\ }\textbf {\bibinfo
  {volume} {61}},\ \bibinfo {pages} {2332} (\bibinfo {year}
  {1988})}\BibitemShut {NoStop}%
\bibitem [{\citenamefont {Higuera}\ and\ \citenamefont
  {Jiménez}(1989)}]{higuera_boltzmann_1989}%
  \BibitemOpen
  \bibfield  {author} {\bibinfo {author} {\bibfnamefont {F.~J.}\ \bibnamefont
  {Higuera}}\ and\ \bibinfo {author} {\bibfnamefont {J.}~\bibnamefont
  {Jiménez}},\ }\href {https://doi.org/10.1209/0295-5075/9/7/009} {\bibfield
  {journal} {\bibinfo  {journal} {Europhysics Letters (EPL)}\ }\textbf
  {\bibinfo {volume} {9}},\ \bibinfo {pages} {663} (\bibinfo {year}
  {1989})}\BibitemShut {NoStop}%
\bibitem [{\citenamefont {Krüger}\ \emph {et~al.}(2017)\citenamefont
  {Krüger}, \citenamefont {Kusumaatmaja}, \citenamefont {Kuzmin},
  \citenamefont {Shardt}, \citenamefont {Silva},\ and\ \citenamefont
  {Viggen}}]{kruger_lattice_2017}%
  \BibitemOpen
  \bibfield  {author} {\bibinfo {author} {\bibfnamefont {T.}~\bibnamefont
  {Krüger}}, \bibinfo {author} {\bibfnamefont {H.}~\bibnamefont
  {Kusumaatmaja}}, \bibinfo {author} {\bibfnamefont {A.}~\bibnamefont
  {Kuzmin}}, \bibinfo {author} {\bibfnamefont {O.}~\bibnamefont {Shardt}},
  \bibinfo {author} {\bibfnamefont {G.}~\bibnamefont {Silva}},\ and\ \bibinfo
  {author} {\bibfnamefont {E.~M.}\ \bibnamefont {Viggen}},\ }\href
  {https://www.springer.com/gp/book/9783319446479} {\emph {\bibinfo {title}
  {The {Lattice} {Boltzmann} {Method}: {Principles} and {Practice}}}},\
  Graduate {Texts} in {Physics}\ (\bibinfo  {publisher} {Springer International
  Publishing},\ \bibinfo {year} {2017})\BibitemShut {NoStop}%
\bibitem [{\citenamefont {Taylor}\ and\ \citenamefont
  {Green}(1936)}]{taylor_mechanism_1936}%
  \BibitemOpen
  \bibfield  {author} {\bibinfo {author} {\bibfnamefont {G.~I.}\ \bibnamefont
  {Taylor}}\ and\ \bibinfo {author} {\bibfnamefont {A.~E.}\ \bibnamefont
  {Green}},\ }\href {https://doi.org/10.1098/rspa.1937.0036} {\bibfield
  {journal} {\bibinfo  {journal} {Proceedings of the Royal Society of London.
  Series A - Mathematical and Physical Sciences}\ }\textbf {\bibinfo {volume}
  {158}},\ \bibinfo {pages} {499} (\bibinfo {year} {1936})}\BibitemShut
  {NoStop}%
\bibitem [{\citenamefont {Dairay}\ \emph {et~al.}(2017)\citenamefont {Dairay},
  \citenamefont {Lamballais}, \citenamefont {Laizet},\ and\ \citenamefont
  {Vassilicos}}]{dairay_numerical_2017}%
  \BibitemOpen
  \bibfield  {author} {\bibinfo {author} {\bibfnamefont {T.}~\bibnamefont
  {Dairay}}, \bibinfo {author} {\bibfnamefont {E.}~\bibnamefont {Lamballais}},
  \bibinfo {author} {\bibfnamefont {S.}~\bibnamefont {Laizet}},\ and\ \bibinfo
  {author} {\bibfnamefont {J.~C.}\ \bibnamefont {Vassilicos}},\ }\href
  {https://doi.org/10.1016/j.jcp.2017.02.035} {\bibfield  {journal} {\bibinfo
  {journal} {Journal of Computational Physics}\ }\textbf {\bibinfo {volume}
  {337}},\ \bibinfo {pages} {252} (\bibinfo {year} {2017})}\BibitemShut
  {NoStop}%
\bibitem [{\citenamefont {Laizet}\ \emph {et~al.}(2019)\citenamefont {Laizet},
  \citenamefont {Lamballais}, \citenamefont {Vassilicos},\ and\ \citenamefont
  {Dairay}}]{laizet_3d_2019}%
  \BibitemOpen
  \bibfield  {author} {\bibinfo {author} {\bibfnamefont {S.}~\bibnamefont
  {Laizet}}, \bibinfo {author} {\bibfnamefont {E.}~\bibnamefont {Lamballais}},
  \bibinfo {author} {\bibfnamefont {J.~C.}\ \bibnamefont {Vassilicos}},\ and\
  \bibinfo {author} {\bibfnamefont {T.}~\bibnamefont {Dairay}},\ }\href
  {https://doi.org/10.5281/zenodo.2577239} {\bibinfo {title} {{3D}
  {Taylor}-{Green} vortex {Direct} {Numerical} {Simulation} statistics from
  {Re}=1250 to {Re}=20000}} (\bibinfo {year} {2019})\BibitemShut {NoStop}%
\bibitem [{\citenamefont {Rogers}\ and\ \citenamefont
  {Moser}(1994)}]{rogers_direct_1994}%
  \BibitemOpen
  \bibfield  {author} {\bibinfo {author} {\bibfnamefont {M.~M.}\ \bibnamefont
  {Rogers}}\ and\ \bibinfo {author} {\bibfnamefont {R.~D.}\ \bibnamefont
  {Moser}},\ }\href {https://doi.org/10.1063/1.868325} {\bibfield  {journal}
  {\bibinfo  {journal} {Physics of Fluids}\ }\textbf {\bibinfo {volume} {6}},\
  \bibinfo {pages} {903} (\bibinfo {year} {1994})}\BibitemShut {NoStop}%
\bibitem [{\citenamefont {Latt}\ \emph {et~al.}(2021)\citenamefont {Latt},
  \citenamefont {Malaspinas}, \citenamefont {Kontaxakis}, \citenamefont
  {Parmigiani}, \citenamefont {Lagrava}, \citenamefont {Brogi}, \citenamefont
  {Belgacem}, \citenamefont {Thorimbert}, \citenamefont {Leclaire},
  \citenamefont {Li}, \citenamefont {Marson}, \citenamefont {Lemus},
  \citenamefont {Kotsalos}, \citenamefont {Conradin}, \citenamefont {Coreixas},
  \citenamefont {Petkantchin}, \citenamefont {Raynaud}, \citenamefont {Beny},\
  and\ \citenamefont {Chopard}}]{latt_palabos_2021}%
  \BibitemOpen
  \bibfield  {author} {\bibinfo {author} {\bibfnamefont {J.}~\bibnamefont
  {Latt}}, \bibinfo {author} {\bibfnamefont {O.}~\bibnamefont {Malaspinas}},
  \bibinfo {author} {\bibfnamefont {D.}~\bibnamefont {Kontaxakis}}, \bibinfo
  {author} {\bibfnamefont {A.}~\bibnamefont {Parmigiani}}, \bibinfo {author}
  {\bibfnamefont {D.}~\bibnamefont {Lagrava}}, \bibinfo {author} {\bibfnamefont
  {F.}~\bibnamefont {Brogi}}, \bibinfo {author} {\bibfnamefont {M.~B.}\
  \bibnamefont {Belgacem}}, \bibinfo {author} {\bibfnamefont {Y.}~\bibnamefont
  {Thorimbert}}, \bibinfo {author} {\bibfnamefont {S.}~\bibnamefont
  {Leclaire}}, \bibinfo {author} {\bibfnamefont {S.}~\bibnamefont {Li}},
  \bibinfo {author} {\bibfnamefont {F.}~\bibnamefont {Marson}}, \bibinfo
  {author} {\bibfnamefont {J.}~\bibnamefont {Lemus}}, \bibinfo {author}
  {\bibfnamefont {C.}~\bibnamefont {Kotsalos}}, \bibinfo {author}
  {\bibfnamefont {R.}~\bibnamefont {Conradin}}, \bibinfo {author}
  {\bibfnamefont {C.}~\bibnamefont {Coreixas}}, \bibinfo {author}
  {\bibfnamefont {R.}~\bibnamefont {Petkantchin}}, \bibinfo {author}
  {\bibfnamefont {F.}~\bibnamefont {Raynaud}}, \bibinfo {author} {\bibfnamefont
  {J.}~\bibnamefont {Beny}},\ and\ \bibinfo {author} {\bibfnamefont
  {B.}~\bibnamefont {Chopard}},\ }\href
  {https://doi.org/10.1016/j.camwa.2020.03.022} {\bibfield  {journal} {\bibinfo
   {journal} {Computers \& Mathematics with Applications}\ }\bibinfo {series}
  {Development and {Application} of {Open}-source {Software} for {Problems}
  with {Numerical} {PDEs}},\ \textbf {\bibinfo {volume} {81}},\ \bibinfo
  {pages} {334} (\bibinfo {year} {2021})}\BibitemShut {NoStop}%
\bibitem [{\citenamefont {Latt}\ and\ \citenamefont
  {Coreixas}(2025)}]{latt_multi-gpu_2025}%
  \BibitemOpen
  \bibfield  {author} {\bibinfo {author} {\bibfnamefont {J.}~\bibnamefont
  {Latt}}\ and\ \bibinfo {author} {\bibfnamefont {C.}~\bibnamefont
  {Coreixas}},\ }\href {https://doi.org/10.48550/arXiv.2506.09242} {\bibinfo
  {title} {Multi-{GPU} {Acceleration} of {PALABOS} {Fluid} {Solver} using {C}++
  {Standard} {Parallelism}}} (\bibinfo {year} {2025})\BibitemShut {NoStop}%
\bibitem [{\citenamefont {Hermite}(1864)}]{hermite1864series}%
  \BibitemOpen
  \bibfield  {author} {\bibinfo {author} {\bibfnamefont {C.}~\bibnamefont
  {Hermite}},\ }\href@noop {} {\bibfield  {journal} {\bibinfo  {journal}
  {Comptes Rendus de l’Académie des Sciences de Paris}\ }\textbf {\bibinfo
  {volume} {52}},\ \bibinfo {pages} {93} (\bibinfo {year} {1864})}\BibitemShut
  {NoStop}%
\bibitem [{\citenamefont {Grad}(1949)}]{grad_note_1949}%
  \BibitemOpen
  \bibfield  {author} {\bibinfo {author} {\bibfnamefont {H.}~\bibnamefont
  {Grad}},\ }\href {https://doi.org/https://doi.org/10.1002/cpa.3160020402}
  {\bibfield  {journal} {\bibinfo  {journal} {Communications on Pure and
  Applied Mathematics}\ }\textbf {\bibinfo {volume} {2}},\ \bibinfo {pages}
  {325} (\bibinfo {year} {1949})}\BibitemShut {NoStop}%
\bibitem [{\citenamefont {Shan}\ \emph {et~al.}(2006)\citenamefont {Shan},
  \citenamefont {Yuan},\ and\ \citenamefont {Chen}}]{shan_kinetic_2006}%
  \BibitemOpen
  \bibfield  {author} {\bibinfo {author} {\bibfnamefont {X.}~\bibnamefont
  {Shan}}, \bibinfo {author} {\bibfnamefont {X.-F.}\ \bibnamefont {Yuan}},\
  and\ \bibinfo {author} {\bibfnamefont {H.}~\bibnamefont {Chen}},\ }\href
  {https://doi.org/10.1017/S0022112005008153} {\bibfield  {journal} {\bibinfo
  {journal} {Journal of Fluid Mechanics}\ }\textbf {\bibinfo {volume} {550}},\
  \bibinfo {pages} {413} (\bibinfo {year} {2006})}\BibitemShut {NoStop}%
\bibitem [{\citenamefont {Malaspinas}(2009)}]{malaspinas_lattice_2009}%
  \BibitemOpen
  \bibfield  {author} {\bibinfo {author} {\bibfnamefont {O.~P.}\ \bibnamefont
  {Malaspinas}},\ }\emph {\bibinfo {title} {Lattice {Boltzmann} method for the
  simulation of viscoelastic fluid flows}},\ \href
  {https://infoscience.epfl.ch/record/140623} {Ph.D. thesis} (\bibinfo {year}
  {2009})\BibitemShut {NoStop}%
\end{thebibliography}
\end{document}